\documentclass[11pt,a4paper]{article}
\pdfoutput=1
\usepackage{jheppub}
\usepackage{rotating}
\usepackage{array}
\usepackage{amsmath}
\usepackage{slashed,hyperref}
\usepackage[normalem]{ulem}
\preprint{DESY 15-036, CERN-PH-TH-2015-057}

\title{Constraining Dark Sectors with Monojets and Dijets}

\author[a]{Mikael Chala,}
\author[a]{Felix Kahlhoefer,}
\author[b]{Matthew McCullough,}
\author[a]{Germano Nardini,}
\author[a]{and \mbox{Kai Schmidt-Hoberg}}

\affiliation[a]{DESY, Notkestrasse 85, D-22607 Hamburg, Germany}
\affiliation[b]{Theory Division, CERN, 1211 Geneva 23, Switzerland}

\emailAdd{mikael.chala@desy.de}
\emailAdd{felix.kahlhoefer@desy.de}
\emailAdd{matthew.mccullough@cern.ch}
\emailAdd{germano.nardini@desy.de}
\emailAdd{kai.schmidt.hoberg@desy.de}

\abstract{We consider dark sector particles (DSPs) that obtain sizeable interactions with Standard Model fermions from a new mediator. While these particles can avoid observation in direct detection experiments, they are strongly constrained by LHC measurements. We demonstrate that there is an important complementarity between searches for DSP production and searches for the mediator itself, in particular bounds on (broad) dijet resonances. This observation is crucial not only in the case where the DSP is all of the dark matter but whenever~--- precisely due to its sizeable interactions with the visible sector~--- the DSP annihilates away so efficiently that it only forms a dark matter subcomponent. To highlight the different roles of DSP direct detection and LHC monojet and dijet searches, as well as perturbativity constraints, we first analyse the exemplary case of an axial-vector mediator and then generalise our results. We find important implications for the interpretation of LHC dark matter searches in terms of simplified models.}


\begin{document}
\maketitle

\section{Introduction}

In the event that the Large Hadron Collider (LHC) discovers evidence for new invisible particles from telltale missing energy signatures, it would not be possible to unequivocally establish if such a dark sector particle (DSP) is a viable dark matter (DM) candidate.  To be able to draw such a conclusion, it is essential to combine LHC searches with cosmological or astrophysical observations as well as with results from direct and indirect detection experiments. Doing so, however, necessarily requires some assumptions on the properties and interactions of the DSP.

By interpreting potential missing energy signatures within the context of a specific theoretical model, it is possible to make concrete comparisons between collider searches, cosmology, and direct and indirect detection signatures~\cite{Beltran:2008xg,Beltran:2010ww}. A well-motivated and simple class of models involves an additional scalar or vector mediator in such a way that the DSP interacts with SM states via the exchange of this new particle. An attractive feature of this framework is that, in the parameter regions currently probed by experiments, the interactions can be large enough to allow for sizeable DM annihilation rates in the early Universe and hence to avoid DM overproduction~\cite{Busoni:2014gta}. One possible example model is a massive $Z'$ arising from a new broken $U(1)_X$ gauge symmetry~\cite{Holdom:1985ag,Babu:1997st}, which can in principle have large couplings to both the DSP and SM fermions~\cite{Dudas:2009uq,Fox:2011qd,Frandsen:2012rk,Alves:2013tqa,Arcadi:2013qia,Lebedev:2014bba,Martin-Lozano:2015vva}.

If this new $s$-channel mediator has a mass comparable to LHC energies, it will affect missing energy signatures at colliders and must therefore be taken into account for the interpretation of DM searches at the LHC and the comparison with direct detection experiments~\cite{Frandsen:2012rk,Buchmueller:2013dya,Buchmueller:2014yoa,Harris:2014hga,Fairbairn:2014aqa,Jacques:2015zha}. Furthermore, the presence of such a mediator has two additional important implications, which motivate the central themes of this work:
\begin{itemize}
\item As well as the usual searches for DM at the LHC based on missing
  energy in association with SM particles one also needs to consider
  dedicated \emph{searches for the mediator particles} themselves,
which make use of the fact that any mediator produced from SM particles in the initial state can also decay back into SM states.
To
  illustrate the complementarity of these two approaches we will
  consider monojet searches~\cite{Khachatryan:2014rra, Aad:2015zva} as a typical representative of the former
  category and compare them to constraints on the mediator arising
  from dijet resonance searches. We will show that, by combining data from
  UA2~\cite{Alitti:1993pn}, the Tevatron~\cite{Aaltonen:2008dn} and the
  LHC~\cite{Khachatryan:2015sja, Aad:2014aqa,
    TheATLAScollaboration:2013gia}, searches for dijet resonances probe a wide range of mediator masses
  from the electroweak scale to well above the TeV scale. Crucially, these searches turn out to be sensitive even for broad
  resonances, thus ruling out large swathes of parameter space which
  would otherwise appear viable when confronted with only monojet and
  direct detection constraints.
\item In the parameter regions probed by LHC monojet and dijet
  searches, the generic prediction is that the DSP would be
  \emph{underproduced in the early Universe}. In particular, if the
  DSP mass $m_\chi$ is larger than the mediator mass $M_R$, the
  process $\chi \bar{\chi} \rightarrow R R$ can easily deplete the DSP
  abundance, and for $m_\chi \sim M_R/2$ annihilation into SM
  fermions will receive a resonant enhancement. 
This prediction is still consistent with all observations if it is assumed that the particle under consideration only constitutes a DM subcomponent and another particle (for example the axion~\cite{Wilczek:1977pj, Weinberg:1977ma}) makes up for the remaining DM abundance.\footnote{An alternative approach would be to call the standard thermal history of the Universe into question and consider additional dilution or production mechanisms for DM~\cite{Gelmini:2010zh}. Such modifications would however require further fields and couplings. In the present paper, we therefore take the standard thermal history as a fundamental underlying assumption.} As a result, the local density of DSPs in the Milky Way can be significantly smaller than what is usually assumed for the interpretation of direct detection experiments. The predicted DSP underabundance must therefore be taken into account when comparing between direct detection and LHC constraints in order to avoid overstating the strength of direct detection limits. This approach leads to a compelling interplay between the different DM detection techniques and will lead us to conclude that the LHC monojets, LHC dijets and direct detection strategies each has a unique foothold in the search for DSPs.
\end{itemize}

\begin{figure}[tb]
\centering
\includegraphics[width=0.95\textwidth]{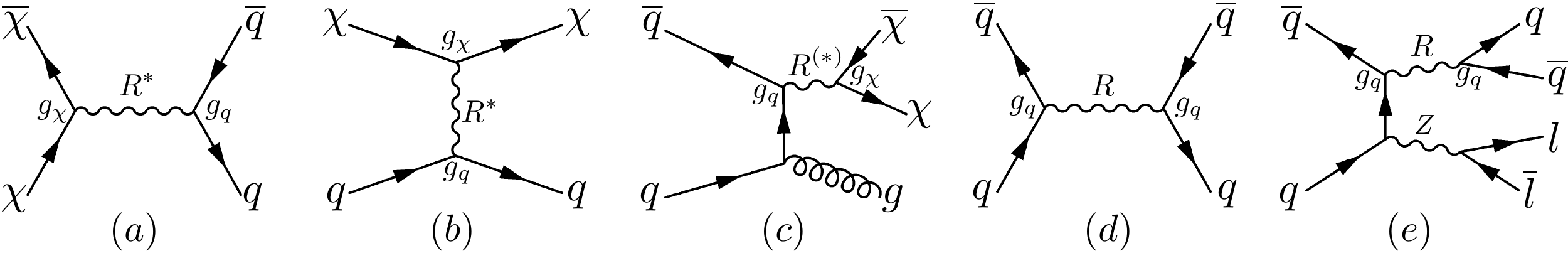}
\caption{The processes considered in this work in terms of visible sector quarks ($q,\overline{q}$), DSPs ($\chi,\overline{\chi}$) and the on-shell (off-shell) mediator particle $R$ ($R^*$).  The various process are: (a)  DM annihilation which sets the relic abundance, (b) DM scattering in direct detection experiments, (c) monojet signatures, in this case due to initial state radiation of a gluon, (d)  LHC Dijet resonance signatures purely through mediator-quark couplings and (e) dijet associated production.}
\label{fig:schem}
\end{figure}

In figure~\ref{fig:schem} we sketch the setup for a dark sector theory involving a DSP $\chi$ and a mediator between the visible sector and the dark sector $R$, together with the detection processes considered in this work. We denote the couplings between the mediator and the visible sector quarks (the DSP) with $g_q$ ($g_\chi$). For the purposes of exploring the broad phenomenology of this dark sector and the general interplay between the different probes let us combine the two couplings into an effective DSP-SM coupling $g = \sqrt{g_q \, g_\chi}$ and consider the effect of varying the coupling $g$. The local density of DSPs in the Milky Way $\rho$ is proportional to the DSP relic abundance from thermal freeze-out $\Omega_\text{DSP}$, which scales as the inverse of the annihilation cross section, i.e.~$\rho \propto \Omega_\text{DSP} \propto g^{-4}$. Any cross section involving interactions between the visible sector and the DSP, such as collider production and direct detection, will scale as $\sigma \propto g^4$~\cite{Cao:2009uw,Beltran:2008xg,Zheng:2010js,MarchRussell:2012hi,Cheung:2012gi} (assuming an off-shell mediator). Thus, broadly speaking, the rate of events in different DM probes have very different scaling with couplings if a standard thermal history is assumed. They are:
\begin{itemize}
\item  Collider searches for missing energy: $\text{Rate} \propto \sigma \propto g^4$\;.
\item  Direct detection: $\text{Rate} \propto (\sigma \times \rho)\propto g^0$\;.
\item  Indirect detection: $\text{Rate} \propto ( \sigma \times \rho^2) \propto g^{-4}$\;.
\end{itemize}
Furthermore, resonance searches at colliders typically depend on the
production cross section for the resonance, $\sigma_R$, multiplied
with the branching ratio into the final state under consideration. If
the (on-shell) mediator has a large branching into light quarks
we hence obtain the final important signature
\begin{itemize}
\item  Collider searches for dijet resonances: $\text{Rate} \propto \sigma_R \propto g^2_{q}$\;.
\end{itemize}
This simple consideration demonstrates that, assuming a standard thermal history and considering the specific phenomenology of the mediator, these four different detection strategies are \emph{parametrically complementary}. In essence, large couplings imply large collider rates but small abundances, small couplings imply small collider rates and large abundances.   Interestingly, to a first approximation, direct detection of a DSP is independent of the interaction strength $g$. Another pertinent consequence is that the dijet constraints break the degeneracy between mediator couplings to the visible and dark sectors.

Of course, the discussion above has been intentionally over-simplified. In reality even in the simplest scenarios there are four relevant parameters, namely $m_\chi$, $M_R$, $g_\chi$ and $g_q$. Specifically, detection strategies often exhibit a highly non-trivial dependence on the mediator mass $M_R$ and also on the ratio $m_\chi / M_R$, which determines whether the mediator may decay to DM, whether there can be resonant enhancements and whether the annihilation channel $\chi \bar{\chi} \rightarrow RR$ is open. In this work we will take all of these parameters into account and furthermore consider different coupling structures for $g_q$ in order to thoroughly explore the relevant phenomenology and interplay between different dark sector probes. 

Moreover, a model containing just a DSP and a mediator is typically only a low-energy effective description of a more complete theory and we therefore have to pay attention to the validity of our approach. In particular, we have to ensure that the model remains perturbative and does not violate unitarity at the energies at which it is being probed~\cite{Griest:1989wd,Shoemaker:2011vi,Busoni:2013lha,Busoni:2014sya,Xiang:2015lfa}. As we will show, these considerations imply that, for the case of an axial-vector mediator, the DM mass must not significantly exceed the mediator mass. The resulting theoretical constraints may not be as concrete as experimental exclusion limits, but must be kept in mind when studying the cosmology and collider phenomenology of a DSP interacting via a mediator.

In the present work we will mostly focus on the scenario of an axial-vector mediator for concreteness and also comment on the case of vector couplings. This choice is motivated by the desire to demonstrate the complementarity between different approaches as clearly as possible as no single probe dominates the experimental limits for this model. Specifically, we will demonstrate that only small, often closed, parameter spaces survive current constraints, providing clear targets for future experimental exploration.
Many of our conclusions apply also to other models of thermal DM that contain a new mediator. The examples chosen here are therefore meant to illustrate how this entire class of dark sector theories can be probed by searches across the frontier of dark sector physics.

The paper is structured as follows. In section~\ref{sec:scenarios} we introduce the general interactions of the DSP and the mediator and discuss the coupling structures and parameter ranges relevant for our study. Section~\ref{sec:relic} then focusses on the calculation of the DSP relic abundance and the resulting rescaling of direct detection bounds. Monojet and dijet constraints will be discussed in section~\ref{sec:monojet} and section~\ref{sec:dijet}, respectively. We present the combination of all of these constraints in section~\ref{sec:combination} and a discussion of more general scenarios and future prospects in section~\ref{sec:discussion}.

\section{Simplified interactions of a vector mediator}
\label{sec:scenarios}

We consider a simplified model of a DSP $\chi$, taken to be a Dirac fermion, and a vector mediator $R$. In this model no gauge symmetry, additional fields to cancel anomalies or symmetry-breaking structure is specified. Thus, as with any effective theory, new dynamics will enter to complete this model in the ultraviolet. We assume that at low scales the specific ultraviolet-completion is decoupled so that $\chi$ and $R$ are the only fields relevant for our study. The couplings of $R$ and $\chi$ are written as
\begin{equation}
\mathcal{L}^{R}_\text{DS} = R_\mu \, \bar{\chi} \, \gamma^\mu (g^V_{\chi}-g^A_{\chi} \gamma^5) \chi \; .
\end{equation}
In the same way, the couplings of $R$ to SM particles are given by
\begin{equation}
\mathcal{L}^{R}_{ f\bar{f}}= \sum_{f=q, \ell}R_\mu \, \bar{f} \, \gamma^\mu (g^V_{f}-g^A_{f}\, \gamma^5)f  \; ,
\end{equation}
where $q$ and $\ell$ denote SM quarks and leptons, respectively. The partial decay widths of the mediator in terms of these couplings are
\begin{align}
\Gamma(R\rightarrow \chi\bar{\chi}) & = \frac{M_R}{12\pi} 
 \sqrt{1-4 \, z_\chi} \, \left[(g^{V}_{\chi})^2+(g^{A}_{\chi})^2 + z_\chi\left(2 (g^{V}_{\chi})^2 - 4 (g^{A}_{\chi})^2\right)  \right] \; , \label{eq:decaywidth} \nonumber \\
\Gamma(R\rightarrow f\bar{f}) & =\frac{M_R \, N_c}{12\pi} 
 \sqrt{1-4 \, z_f} \, \left[(g^{V}_{f})^2+(g^{A}_{f})^2 + z_f\left(2 (g^{V}_{f})^2 - 4 (g^{A}_{f})^2\right)  \right] \; ,
\end{align}
where $z_{\chi,f} = m_{\chi,f}^2/M_R^2$, $N_c = 3$ for quarks and $N_c = 1$ for leptons.

If both $g^V_\chi$ and $g^V_q$ are non-zero, the DSP interacts with nuclei via spin-independent (SI) scattering. These interactions receive a coherent enhancement proportional to the square of the target nucleus mass and are therefore strongly constrained by direct detection experiments~\cite{Akerib:2013tjd} (see e.g.~\cite{Alves:2015pea} for a recent discussion of these constraints in the context of vector mediators). In this case, LHC bounds typically give no relevant constraints unless the DSP is very light.

For the largest part of this work we will therefore assume that at least one of the vector couplings vanishes.\footnote{For example, the vector coupling $g^V_\chi$ automatically vanishes if the DSP is a Majorana fermion.} In this case, DSP-nucleus scattering will be dominated by the spin-dependent (SD) interactions induced by the axial couplings. Potential cross-terms such as $g^V_q \, g^A_\chi$ lead to momentum suppressed scattering in the non-relativistic limit, which can safely be neglected.\footnote{Note that these interactions also induce SI but mass-suppressed scattering at one-loop level. The resulting contributions are, however, subdominant to tree-level SD interactions~\cite{Haisch:2013uaa,Crivellin:2014qxa,D'Eramo:2014aba}.} To simplify our analysis, we will set both $g^V_q$ and $g^V_\chi$ equal to zero unless explicitly stated otherwise.
More general couplings structures will be discussed in section~\ref{sec:vector}. 

We assume that there is no direct link between the quark couplings $g_q$ and the leptonic couplings $g_\ell$, in contrast to the case in which the mediator $R$ obtains its SM couplings from mixing with the $Z$ boson. The reason is that the leptonic couplings are very tightly constrained by searches for dilepton resonances~\cite{Dudas:2009uq,Arcadi:2013qia, Lebedev:2014bba}. Rather than forcing the quark couplings to be equally small, we want to consider the case where the mediator couples much more strongly to quarks than to leptons (see e.g.~\cite{Duerr:2013lka, Duerr:2014wra} for a discussion of a baryonic $Z'$). In this case, the leptonic couplings will not give a relevant contribution to the DM phenomenology of the model and we can simply set $g^V_\ell = g^A_\ell =0$. Moreover, it was shown in~\cite{Frandsen:2012rk} that similarly stringent constraints apply to the interactions between $R$ and SM gauge bosons as well as the Higgs boson, which are therefore also assumed to be negligible.

In summary, we will focus on the following simplified interactions:
\begin{equation}
 \mathcal{L} \supset g^A_\chi \, \bar{\chi} \gamma^\mu \gamma^5 \chi \, R_\mu + \sum_q g^A_q \, \bar{q} \gamma^\mu \gamma^5 q \, R_\mu \; .
\end{equation}
We allow for the possibility to have different couplings for the individual quark flavours. To study how the phenomenology of the model depends on the different couplings, we will consider a number of different scenarios:
\begin{itemize}
 \item \emph{Universal couplings}: $g^A_u = g^A_d = g^A_s = g^A_c = g^A_b = g^A_t \equiv g^A_q$.
 \item \emph{Isovector couplings}: $g^A_u = g^A_c = g^A_t = - g^A_d = -g^A_s = -g^A_b \equiv g^A_q$.
 \item \emph{No heavy-quark couplings}: $g^A_u = g^A_d = g^A_s = g^A_c \equiv g^A_q$ and $g^A_b = g^A_t = 0$.
\end{itemize}
As we will see below, the heavy-quark couplings play a non-negligible role, because they can provide additional annihilation channels for the DSP and thereby reduce the total DSP abundance (thus weakening bounds from direct detection experiments) and because they can increase the width of the mediator (thus weakening bounds from LHC searches).

Clearly, there is no fundamental reason why $g^A_\chi$ should be identical to $g^A_q$. Even if both couplings arise from a new gauge group, the respective charges and hence the resulting couplings could easily differ by a factor of a few~\cite{An:2012va, An:2012ue}. Alternatively, if the DSP couples directly to the mediator and couplings to quarks are generated only through a small mixing, $g^A_\chi$ could easily be significantly larger than $g^A_q$. In the following, we will usually consider the three cases $g^A_\chi = g^A_q $, $g^A_\chi = 4 \, g^A_q$ and $g^A_\chi = 9 \, g^A_q$.

\begin{figure}[tb]
\centering
\includegraphics[width=0.49\textwidth]{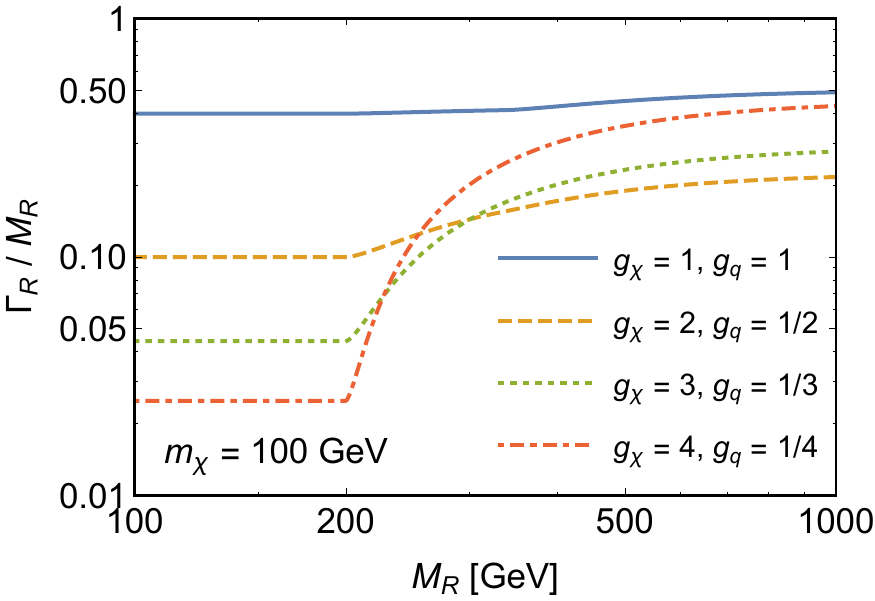}\quad
\includegraphics[width=0.47\textwidth,clip,trim=0 -4 0 0]{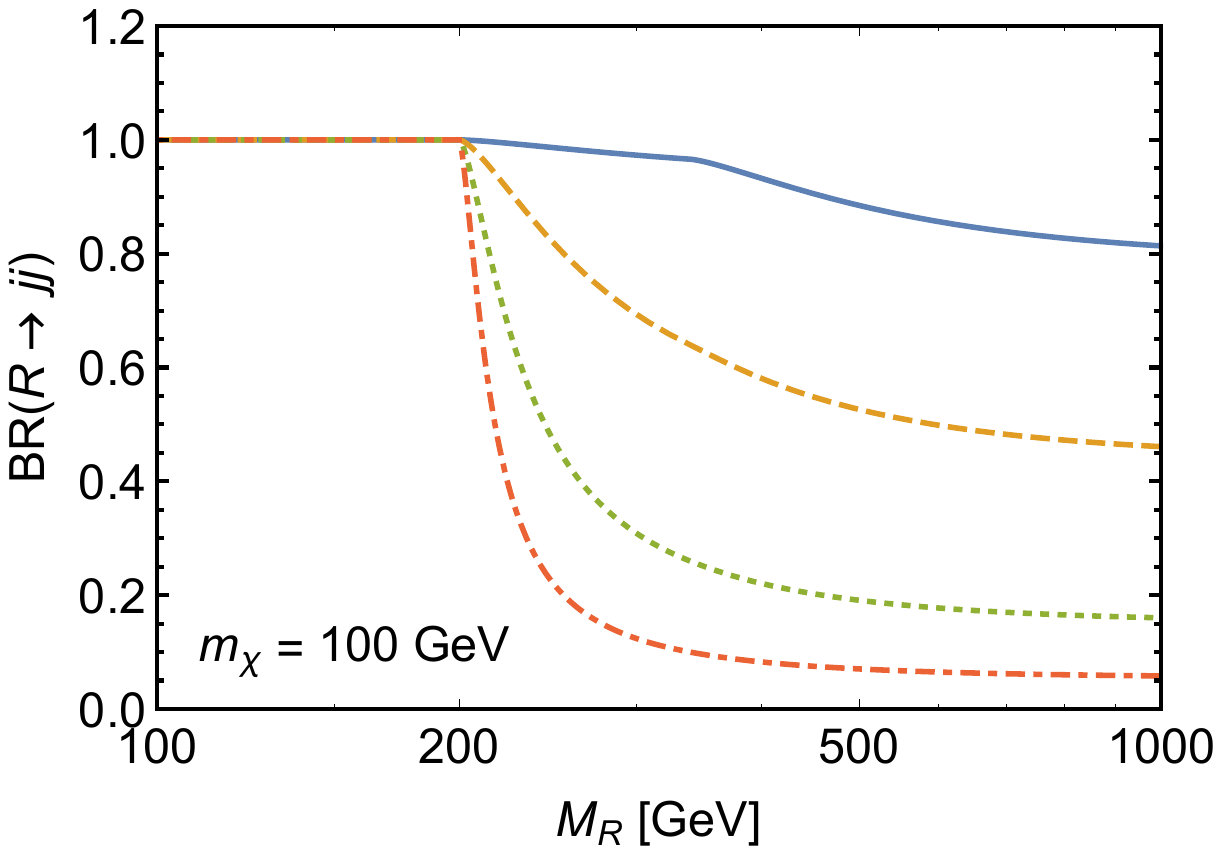}
\caption{Left: Total width of the mediator $R$ divided by $M_R$ as a function of the mediator mass for various coupling ratios $g^A_\chi / g^A_q$. Right: The corresponding branching ratio into dijets as a function of $M_R$. For both plots, the DSP mass has been fixed to $m_\chi = 100\:\text{GeV}$.}
\label{fig:widths}
\end{figure}

Since for processes involving an off-shell mediator, cross sections are typically proportional to $(g^A_q)^2 \, (g^A_\chi)^2$, it will be useful to consider the effective coupling $g \equiv (g^A_q \, g^A_\chi)^{1/2}$. In figure~\ref{fig:widths} we show the total width of the mediator, as well as the corresponding branching ratios into jets, as a function of $M_R$ for various coupling ratios $g^A_\chi / g^A_q$. For $M_R \gg m_\chi$ we find that the mediator decays dominantly into quarks if $g^A_\chi / g^A_q \lesssim 4$ and dominantly into the DSP if $g^A_\chi / g^A_q \gtrsim 4$. If the coupling ratio is approximately $g^A_\chi / g^A_q \sim 4$ both decay channels have comparable branching ratios, leading to the smallest mediator width for fixed $g$.

To conclude this section, let us briefly discuss the parameter ranges of interest. First of all, the magnitude of the couplings is bounded from above by the requirement of perturbativity, which implies $g^A_{q,\chi} < \sqrt{4 \pi}$. In addition, we require that the mediator width satisfies $\Gamma_R / M_R < 1$, such that it makes sense to interpret the mediator as a new particle. This second requirement implies roughly $g^A_{q} \lesssim 1.5$ and $g^A_{\chi} \lesssim 6$. In practice, we will focus on the two cases $g = 1$ and $g = 0.5$, which for the coupling ratios discussed above give $\Gamma_R / M_R < 0.5$. We do not consider the case $\Gamma_R / M_R > 0.5$, because such mediators would typically be very difficult to observe in searches for dijet resonances.

For the mediator mass, we restrict ourselves to the case $M_R > 100\:\text{GeV}$. The reason is that we want to avoid the case where the mediator mass is very close to the mass of the $Z$-boson $M_Z$, because for $M_R \approx M_Z$ quark loops induce large mixing between the two gauge bosons, leading to conflicts with electroweak precision observables~\cite{Carone:1995pu,Babu:1997st,Chun:2010ve,Frandsen:2011cg}. Parametrically, the kinetic mixing $\mathcal{L}\supset \epsilon \, \partial^\mu Z^\nu (\partial_\mu R_\nu - \partial_\nu R_\mu)$ is given by
\begin{equation}
\epsilon \sim \sum_q \frac{(g^A_q)^2}{16 \pi^2} \sim 10^{-2} (g^A_q)^2
\end{equation}
and the correction to the $\rho$ parameter, $\Delta \rho \equiv M_W^2 / (M_Z^2 \, \cos^2 \theta_\text{W}) - 1$,  is roughly
\begin{equation}
\Delta\rho \sim \epsilon^2 \frac{M_Z^2}{M_R^2 - M_Z^2} \; ,
\end{equation}
which is sufficiently small for $g^A_q \lesssim 1$ and $M_R \gtrsim 100\:\text{GeV}$.\footnote{Another concern are loop-induced couplings of the mediator to leptons via the $Z$-mixing, leading to potential signals in searches for dilepton resonances. We have checked that these signals give weaker constraints than searches for dijet resonances.}

Finally, if the theory is chiral, i.e.\ if $g^A_\chi \neq 0$, the DSP mass cannot be raised arbitrarily compared to the mediator mass. The reason is that in a chiral gauge theory we essentially have to invoke a Higgs mechanism to generate both the DSP mass and the mediator mass:
\begin{equation}
\mathcal{L} \supset - \left(D^\mu \Phi\right)^\dagger D_\mu \Phi - (y_\chi \, \Phi \, \bar{\chi}_L \chi_R + \text{h.c.}) \; ,
\end{equation}
where $\Phi$ is the Higgs field in the dark sector and $D^\mu = \partial^\mu - i g' q \, R^\mu$ with $g'$ and $q$ being the gauge coupling of the new $U(1)'$ and the corresponding charge. In order for the Yukawa interaction to be gauge invariant, we must require $q_\Phi = q_{\chi_L} - q_{\chi_R}$.
Once $\Phi$ aquires a vacuum expectation value, $\langle \Phi \rangle = v'/\sqrt{2}$, we obtain the mass terms
\begin{equation}
\mathcal{L} \supset - \frac{1}{2} g'^2 \, q_\Phi^2  \, v'^2 \, R^\mu R_\mu - \frac{1}{\sqrt{2}} y_\chi \, v' \, \bar{\chi} \chi\; ,
\end{equation}
implying $M_R = g' \, q_\Phi \, v' = g' \, (q_{\chi_L} - q_{\chi_R}) \, v' = g^A_\chi \, v'$ and $m_\chi = y_\chi \, v'/\sqrt{2} = y_\chi \, M_R / (\sqrt{2} \, g_\chi^A)$. 
In order for the Yukawa interaction to remain perturbative we impose the bound
\begin{equation}
 m_\chi \lesssim \frac{\sqrt{4\pi}}{g^A_\chi} M_R \; .
 \label{eq:pertYuk}
\end{equation}
We note that, even if the mediator mass and the DSP mass are generated in a different way, a similar constraint must apply to ensure that the coupling between the longitudinal component of $R$ and the DSP remains perturbative. Indeed, we will see below that this inequality is closely related to the requirement of perturbative unitarity in DSP scattering and annihilation.

\section{Non-collider constraints}
\label{sec:relic}
\subsection{Relic density}

For the Lagrangian introduced above, the DSP annihilation cross section into quarks is
\begin{align}
\sigma v(\chi \bar{\chi} \rightarrow q \bar{q}) \simeq \frac{3 \, m_\chi^2 \, \sqrt{1-z_q}}{2 \pi \left[(M_R^2 - 4 m_\chi^2)^2 + (\Gamma_R \, M_R)^2\right]} \bigg(
& (g^V_\chi)^2 \left[(g^V_q)^2 (2 + z_q) + 
2 \, (g^A_q)^2 (1 - z_q) \right] \nonumber \\ & +
(g^A_q)^2 (g^A_\chi)^2 z_q \frac{(4 m_\chi^2 - M_R^2)^2}{M_R^4} \bigg) \; ,
\label{eq:ann}
\end{align}
where $z_q = m_q^2 / m_\chi^2$ and we have chosen unitary gauge for the propagator of the vector boson as this should capture the effects of the Goldstone bosons present in an ultraviolet-complete realisation of this model.

If $g^V_\chi$ is zero, the annihilation cross section is proportional to $m_f^2 / m_\chi^2$, i.e.\ there is a helicity suppression for annihilation into light quarks. In this case, it is important to also include the $p$-wave contribution for calculating the DSP relic abundance. Expanding $\sigma v = a + b v^2$, we find (setting $g^V_q = g^V_\chi = 0$)
\begin{equation}
b = \frac{(g^A_\chi)^2(g^A_q)^2 m_\chi^2}{2 \pi \left[(M_R^2 - 4 m_\chi^2)^2 + (\Gamma_R \, M_R)^2\right]} (1 - z_f)^{3/2} \;. 
\end{equation}

For $m_\chi > M_R$ the DSP can also annihilate directly into the mediator, which then subsequently decays into SM particles. The corresponding annihilation cross section is given by
\begin{align}
 \sigma v(\chi \bar{\chi} \rightarrow R R) = & \frac{(m_\chi^2 - M_R^2)^{3/2}}{4 \pi \, m_\chi M_R^2 (M_R^2 - 2 m_\chi^2)^2} \nonumber \\ & \times \Bigl( 8 (g^A_\chi)^2 (g^V_\chi)^2 m_\chi^2 + \left[ (g^A_\chi)^4 - 6 (g^A_\chi)^2 (g^V_\chi)^2 + 
      (g^V_\chi)^4 \right] M_R^2 \Bigr) \; .
\end{align}

\begin{figure}[tb]
\centering
\includegraphics[width=0.45\textwidth]{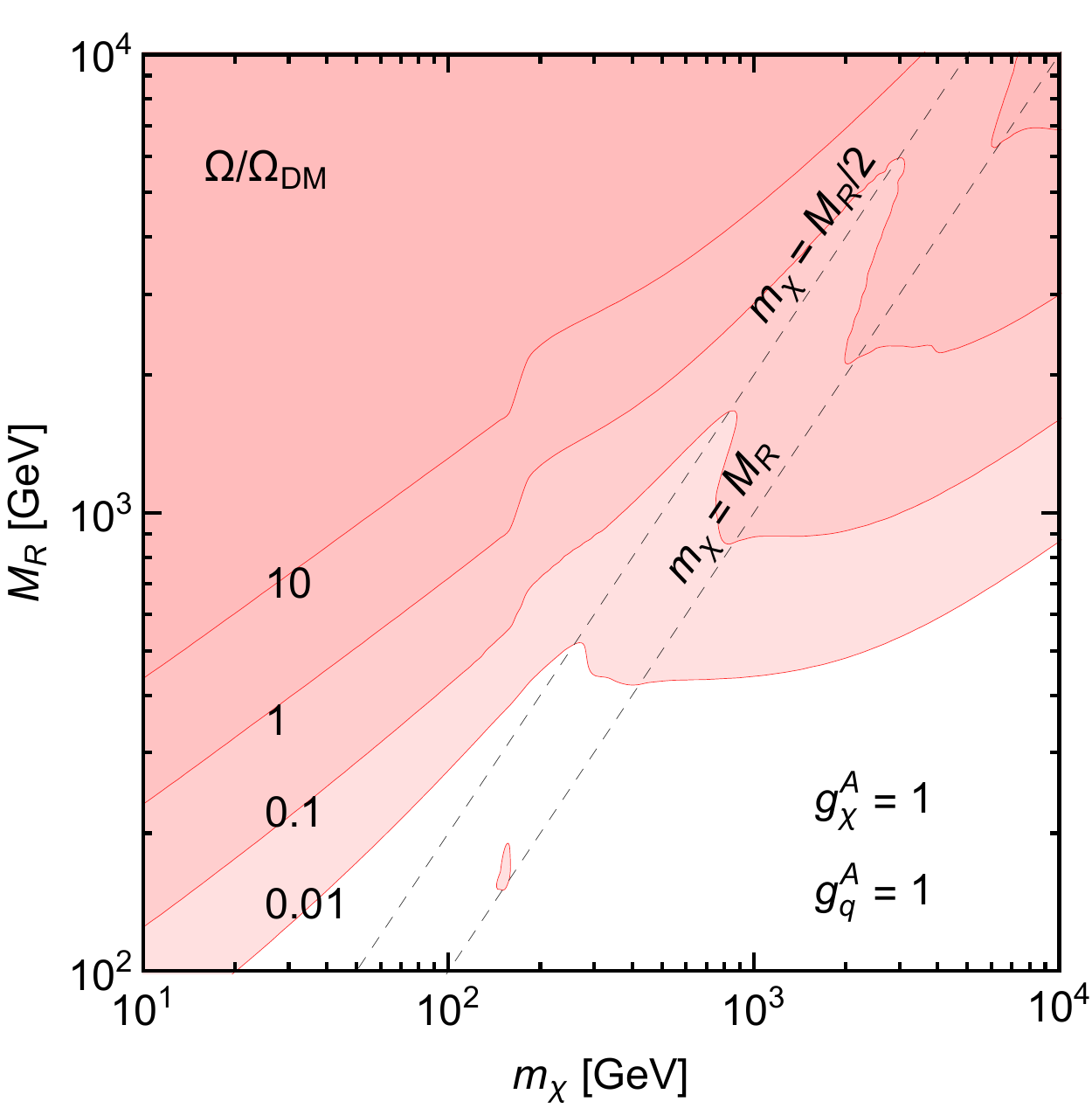}\quad
\includegraphics[width=0.45\textwidth]{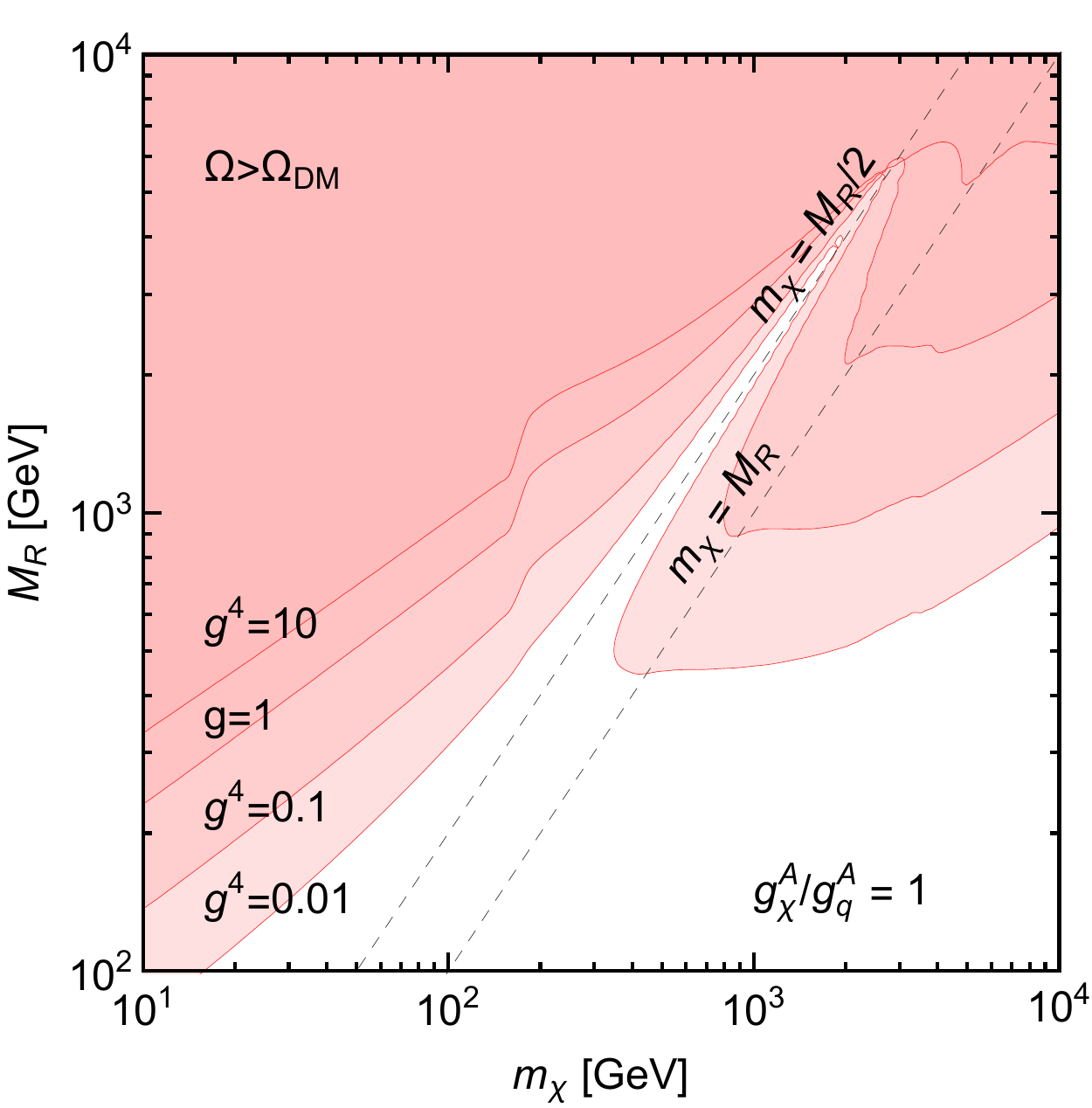}
\caption{The relic abundance $\Omega_\text{DSP}$ of the DSP as a function of $m_\chi$ and $M_R$. Left panel: Contours of constant $\Omega_\text{DSP}$ for $g^A_\chi = g^A_q = 1$. Right panel: The parameters giving $\Omega_\text{DSP} = \Omega_\text{DM}$ for different choices of couplings. The two plots are not equivalent since the mediator width and the freeze-out temperature $T_\text{f}$ depend on the couplings.}
\label{fig:relic}
\end{figure}

To calculate the relic density of the DSP, we have implemented both types of interactions in \texttt{micrOMEGAs v3}~\cite{Belanger:2013oya}, which numerically solves the Boltzmann equation.
As \texttt{micrOMEGAs} takes into account the full expressions for the annihilation cross section (rather than the velocity expansions shown above), the calculation gives reliable results also close to resonance. We show the results of this calculation in figure~\ref{fig:relic}. The left panel shows contours of constant relic density for fixed couplings $g^A_\chi = g^A_q = 1$. The right panel shows the parameter region excluded by the requirement $\Omega_\text{DSP} \leq \Omega_\text{DM} \approx 0.119 / h^2$ for different choices of the coupling product $g \equiv (g^A_q \, g^A_\chi)^{1/2}$, keeping the coupling ratio fixed to $g^A_\chi / g^A_q = 1$. In the approximation that $\Omega_\text{DSP}$ is proportional to $g^{-4}$ this plot contains the same information as the one on the right-hand side. For example, the line $\Omega_\text{DSP} = 10^{-1} \, \Omega_\text{DM}$ for $g=1$ in the left panel corresponds to the exclusion line $\Omega_\text{DSP} = \Omega_\text{DM}$ for $g^4 = 0.1$ in the right panel. Visible differences arise, however, from the fact that $\Omega_\text{DSP}$ depends on the mediator width, as well as on the freeze-out temperature, which can be significantly different from the usual choice $T_\text{f} \approx m_\chi / 20$ in the case that $\Omega_\text{DSP}$ is very different from $\Omega_\text{DM}$.

We would like to emphasise that~--- strictly speaking~--- the calculation performed above provides only an upper limit on the relic abundance of the DSP, which could be further reduced in the presence of additional annihilation channels. As argued in section~\ref{sec:scenarios}, however, experimental constraints essentially require the DSP to couple much more weakly to leptons and to SM bosons than to quarks, so that it appears difficult to obtain sizeable contributions from any of these interactions. In any case, such additional annihilation channels would only reduce the abundance of the DSP and therefore further weaken direct detection and indirect detection compared to LHC searches, so that we essentially consider the case that is most optimistic for direct detection.\footnote{It is rather difficult in general to raise the DSP abundance above the value predicted by naive thermal freeze-out. One attractive possibility would be to assume an initial particle-antiparticle asymmetry in the dark sector. In this case, the DSP relic abundance is essentially set by this asymmetry and does therefore not directly reflect the interaction strength of the DSP. We will return to this possibility in section~\ref{sec:discussion}.}

Before we turn to the discussion of direct detection constraints, let us briefly return to eq.~(\ref{eq:ann}). Unitarity arguments require that the cross section behaves as $\sigma v \propto 1/s$ at high energies, which in the case of DM annihilation implies $\sigma v \propto 1/m_\chi^2$.  Thus in eq.~(\ref{eq:ann}) the term proportional to the axial couplings appears to violate unitarity in the limit $m_\chi \rightarrow \infty$ as the cross section scaling becomes $\sigma v \propto m_\chi^2/M_R^4$. As discussed above, however, this limit is not physical, due to the diverging couplings between the DSP and the longitudinal component of $R$.  Imposing the perturbativity constraint from eq.~(\ref{eq:pertYuk}), rearranged to $M_R > g^A_\chi \, m_\chi/ \sqrt{4\pi}$, is sufficient to make sure that the annihilation cross section remains well-behaved with $\sigma v \propto 1/m_\chi^2$ for large DM mass, thus we will impose this constraint on the parameter space throughout. It would be interesting to investigate perturbative unitarity in this model further to see if this constraint could be strengthened.

\subsection{Direct detection}

As noted above direct detection experiments constrain the rate of interactions between the DSP and nuclei, which is proportional to the local DM
density $\rho$. Using the results shown in figure~\ref{fig:relic}, we can therefore now rescale the bounds from direct detection experiments under the assumption that $\rho$ is proportional to $\Omega_\text{DSP}$.\footnote{This is a good approximation in the case that all of the DM is cold and collisionless. If, on the other hand, the dark sector contains a mix of warm and cold DM or contributions from DSPs with large self-interactions or dissipation, the scaling can become more complicated. We do not consider this case further.} To obtain constraints we use the recent results from the LUX experiment~\cite{Akerib:2013tjd}. 

The SI and SD DSP-nucleon scattering cross sections at zero momentum transfer are given by
\begin{equation}
\sigma^\text{SI}_N = f_N^2 \frac{\mu^2_{N\chi}}{\pi \, M_R^4}
\label{eq:DD}
\end{equation}
and
\begin{equation}
\sigma^\mathrm{SD}_N =  a_N^2 \frac{3 \, \mu^2_{N\chi}}{\pi \, M_R^4} \;,
\label{eq:SDaxial}
\end{equation}
where $N$ stands for either $p$ or $n$ and $\mu_{N \chi} = m_\chi \, m_N / (m_\chi + m_N)$ is the reduced DSP-nucleon mass. The effective DSP-nucleon couplings for SI interactions are given by $f_p = g^\mathrm{V}_\chi (2 g^\mathrm{V}_u + g^\mathrm{V}_d)$ and $f_n = g^\mathrm{V}_\chi (g^\mathrm{V}_u + 2 g^\mathrm{V}_d)$. For SD scattering, we can write
\begin{align}
a_{p,n} = g^\mathrm{A}_\chi \sum_{q=u,d,s} \Delta q^{(p, n)} \, g^\mathrm{A}_q \; .
\end{align}
The coefficients $\Delta q^{(N)}$ encode the contributions of the light quarks to the nucleon spin, which can be extracted from polarised deep inelastic scattering. The Particle Data Group values are~\cite{Beringer:1900zz}
\begin{align}
\Delta u^{(p)} & = \Delta d^{(n)} = 0.84 \pm 0.02 \; , \nonumber \\
\Delta d^{(p)} & = \Delta u^{(n)} = -0.43 \pm 0.02 \; , \label{eq:deltas} \\
\Delta s^{(p)} & = \Delta s^{(n)} = -0.09 \pm 0.02 \; . \nonumber
\label{eq:Delta}
\end{align}

\begin{figure}[tb]
\centering
\includegraphics[width=0.45\textwidth]{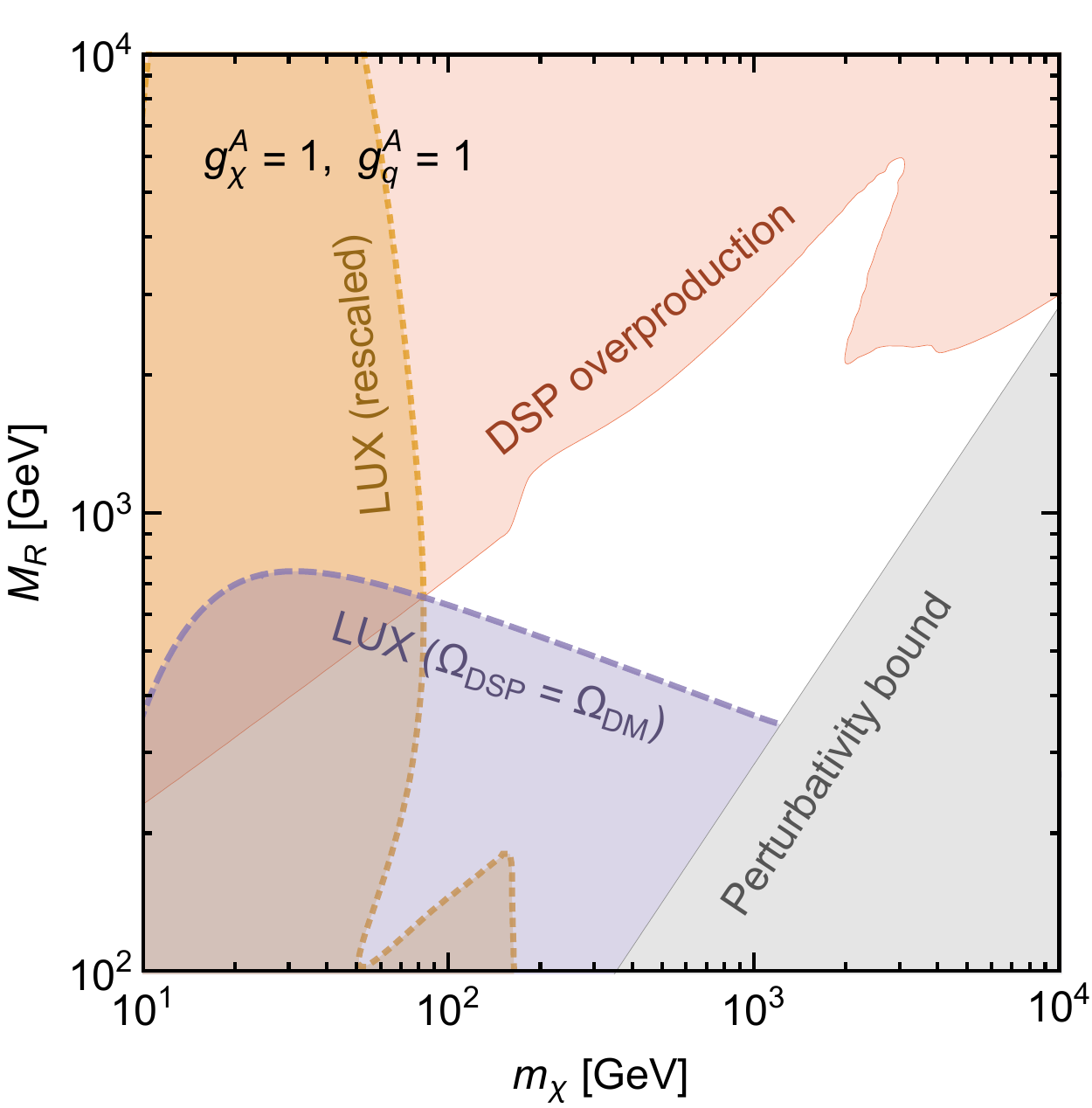}\quad
\includegraphics[width=0.45\textwidth]{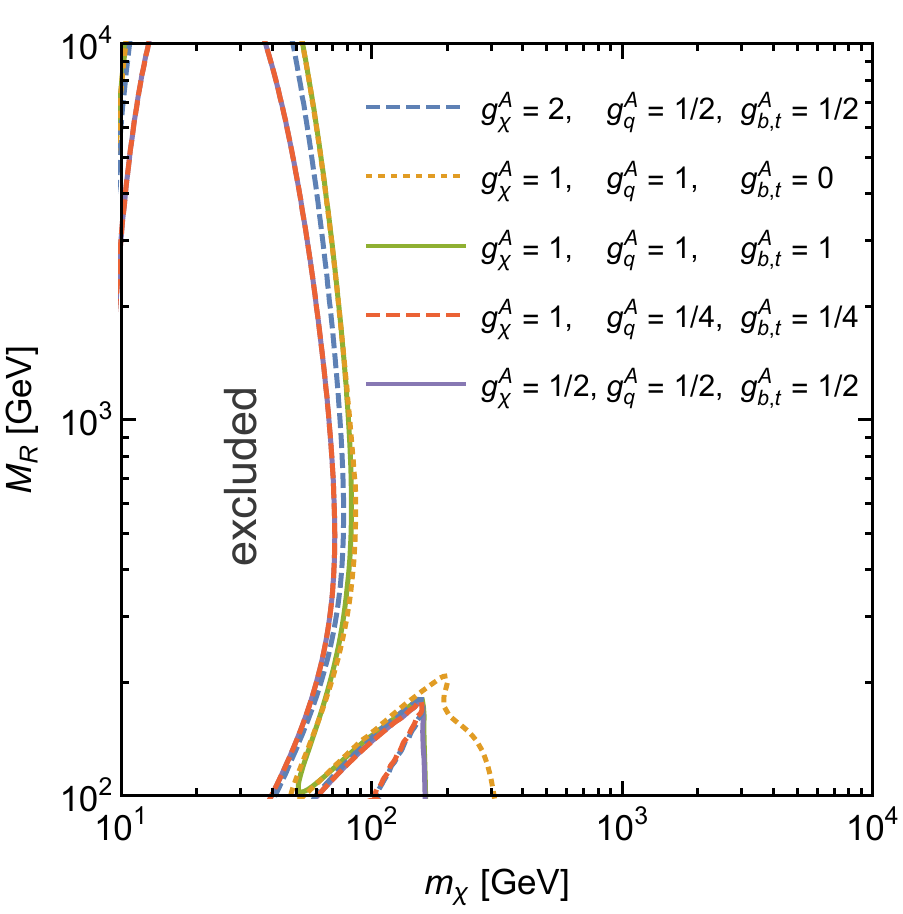}
\caption{Parameter regions excluded by LUX. Left panel: The conventional LUX bound for $\Omega_\text{DSP}$ fixed to $\Omega_\text{DM}$ for $g^A_q = g^A_\chi = 1$ (purple) compared to the relic density constraint (red) and the rescaled LUX bound (orange). The perturbativity bound from eq.~\eqref{eq:pertYuk} is shown in grey. Right panel: The rescaled LUX bound for different coupling choices.}
\label{fig:DD}
\end{figure}

For most of the relevant parameter space the strongest constraints on SI and SD interactions come from the recent LUX experiment~\cite{Akerib:2013tjd}. However, a dedicated LUX analysis for SD constraints has not yet been performed and thus we calculate our own constraints, similarly to~\cite{Buchmueller:2014yoa,Savage:2015xta, Martin-Lozano:2015vva}. We use the spin structure functions from \cite{Klos:2013rwa} and the same limit-setting procedure as described in~\cite{Fox:2014kua} to determine the parameter region excluded at 95\% confidence level (C.L.). This procedure leads to good agreement with the published LUX bound for the case of SI interactions in the DM mass range that we consider.

The resulting constraints are shown in figure~\ref{fig:DD}. The left
panel illustrates how the direct
detection bound is constructed. The grey region indicates the parameter space where the perturbativity bound from eq.~\eqref{eq:pertYuk} is violated. The purple shaded region
 shows the parameter space conventionally
excluded by LUX when assuming $\Omega = \Omega_\text{DM}$ for $g^A_q =
g^A_\chi = 1$. The red shaded region shows the parameter region
excluded by the requirement not to overproduce the DSP. Finally, the
orange region shows the parameter region
excluded by the rescaled direct detection bound (allowing both
$\Omega_\text{DSP} < \Omega_\text{DM}$ and $\Omega_\text{DSP} >
\Omega_\text{DM}$) obtained by taking the thermal relic density as a
prediction of the theory. Clearly, in the red shaded region (where
$\Omega_\text{DSP}$ is larger than $\Omega_\text{DM}$) the rescaled
bound is stronger than the conventional bound, while outside of this
region it is weaker because the DSP constitutes only a fraction of the
DM abundance. By construction, both bounds coincide along the line
$\Omega_\text{DSP} = \Omega_\text{DM}$.

The right panel of figure~\ref{fig:DD} shows the rescaled direct detection bounds for a variety of different couplings. As expected, we find that (for fixed ratio of $g^A_\chi$ and $g^A_q$) the bounds are largely invariant under an overall rescaling of the couplings, because the change in the scattering cross section compensates the change in the DSP density. Nevertheless, there are small observable differences due to non-negligible changes in the mediator width and the freeze-out temperature. The largest changes are observed in the parameter region $m_\chi > M_R, \, m_t$, where the annihilation channels $\chi \bar{\chi} \rightarrow t \bar{t}$ and $\chi \bar{\chi} \rightarrow R R$ open up and introduce a dependence of the relic density on $g^A_{b,t}$ and the coupling ratio $g^A_\chi / g^A_q$. In summary, we see that by taking the reasonable assumption that the thermal abundance is a prediction of the model the constraints from direct detection on a particular theory are substantially altered, and much larger regions of parameter space are allowed than if the assumption of $\Omega = \Omega_\text{DM}$ is imposed.

\subsection{Indirect detection}

Finally, we would like to point out that in the parameter region allowed by the requirement $\Omega_\text{DSP} \leq \Omega_\text{DM}$ there are typically no observable signals from DM indirect detection in any present experiment and in particular our model does not provide an explanation for the diffuse GeV-energy excess of gamma-ray emission from the Galactic Centre observed with the Fermi-LAT instrument~\cite{Goodenough:2009gk,Hooper:2010mq,Hooper:2011ti,Abazajian:2012pn,Daylan:2014rsa}. The reason is that for $m_\chi < m_t$ the $s$-wave contribution to the DM annihilation cross section is helicity suppressed, so that freeze-out is dominated by the $p$-wave contribution, which becomes unobservably small in the present Universe. As a consequence, the parameter region in which the DSP corresponds to all of DM corresponds to $\sigma v \ll 2.5 \cdot 10^{-26} \text{cm}^3/\text{s}$ in the present Universe. For $m_\chi > m_t$, on the other hand, there are presently no indirect detection experiments sensitive to the thermal cross section $\sigma v \approx 2.5 \cdot 10^{-26} \text{cm}^3/\text{s}$ for $\Omega_\text{DSP} = \Omega_\text{DM}$. 

Of course, it is possible in our model to obtain significantly larger annihilation cross sections, but only at the expense of depleting the DSP abundance. Since annihilation signals depend on the square of the DSP density, making the annihilation cross section larger will reduce rather than enhance the magnitude of any indirect detection signal. Consequently, one can never exceed the signal strength expected in the case that the DSP constitutes all of the DM.\footnote{Annihilation of DM subcomponents can however provide a plausible explanation for indirect detection signatures corresponding to annihilation cross sections below the thermal one.}

\section{Monojet searches}
\label{sec:monojet}

LHC searches for jets in association with missing transverse energy ($\slashed{E}_T$) place strong constraints on the interactions between quarks and the DSP~\cite{Khachatryan:2014rra, Aad:2015zva}. These constraints are most easily interpreted in terms of contact interactions~\cite{Cao:2009uw, Beltran:2010ww,Bai:2010hh,Goodman:2010ku,Zheng:2010js,Rajaraman:2011wf,Goodman:2010yf,Fox:2011pm,Shoemaker:2011vi,Cheung:2012gi,Dreiner:2013vla,Busoni:2013lha,Busoni:2014sya,Racco:2015dxa}, but have also been interpreted in terms of the exchange of a vector mediator~\cite{Frandsen:2012rk,Fox:2012ru,Buchmueller:2013dya,Buchmueller:2014yoa,Harris:2014hga,Fairbairn:2014aqa,Jacques:2015zha}. Here we follow closely the analysis presented in~\cite{Buchmueller:2014yoa} and find good agreement with the results presented there.

If the mediator is forced to be off-shell in the production of the DSP (either because it is too heavy to be produced on-shell at the LHC or because $M_R < 2 m_\chi$ so that decays into the DSP are kinematically forbidden), the monojet cross section at the LHC depends on the couplings of the mediator according to $\sigma(pp \rightarrow j \chi \bar{\chi} ) \propto g^4$. If the mediator can be produced on-shell (and provided that we can treat the resonance in the narrow-width approximation) the monojet cross section will be proportional to $\sigma(pp \rightarrow j \chi \bar{\chi} ) \propto (g^A_q)^2 \times \text{BR}(R \rightarrow \chi \bar{\chi})$. As long as we keep the coupling ratio $g^A_\chi / g^A_q$ fixed, the branching ratios of the mediator are independent of $g$, leading to $\sigma(pp \rightarrow j \chi \bar{\chi} ) \propto g^2$.

In many realistic cases, however, the width of the mediator may become so large that the narrow-width approximation is no longer valid. Furthermore, for $m_\chi$ close to $M_R / 2$ there can be relevant contributions from both on-shell and off-shell mediators. In practice, monojet cross sections can therefore depend on all relevant parameters ($m_\chi$, $M_R$, $g^A_\chi$ and $g^A_q$) in a non-trivial way.

To derive exclusion limits, we consider the most recent monojet search from CMS at 8 TeV, based on an integrated luminosity of $19.7\:\text{fb}^{-1}$~\cite{Khachatryan:2014rra}. This CMS analysis considers events with large amounts of $\slashed{E}_T$ provided there is a primary jet ($j_1$) with transverse momentum $p_T>110\:\text{GeV}$ and pseudorapidity $|\eta|<2.4$. A secondary jet ($j_2$) with $p_T>30\:\text{GeV}$ is also permitted if the two jets are not back-to-back: $|\Delta\phi(j_1, j_2)|<2.5$. Events with high-$p_T$ tertiary jets, electrons or muons are vetoed. The analysis considers a number of different requirements for the amount of $\slashed{E}_T$ in the range of 250--$550\:\text{GeV}$.

We simulate monojet events using the implementation of an axial-vector mediator in the \texttt{Powheg-Box v2}~\cite{Haisch:2013ata}. We employ the \texttt{MSTW2008} parton distribution functions (PDFs)~\cite{Martin:2009iq} and set the renormalisation and factorisation scale $\mu$ dynamically, choosing $\mu = H_T / 2$ with $H_T = \sqrt{m_{\chi \bar{\chi}}^2 + p_{T,j_1}^2} + p_{T,j_1}$. For showering and hadronisation we use \texttt{Pythia v6}~\cite{Sjostrand:2006za}. It was shown in~\cite{Haisch:2013ata} that~--- in the presence of a veto on tertiary jets~--- including next-to-leading order corrections in combination with parton showering only leads to a modest change in the monojet cross section of less than 10\%. We neglect this small enhancement here and generate all events at leading order.

\begin{figure}[tb]
\centering
\includegraphics[width=0.97\textwidth,clip,trim=-178 0 0 0]{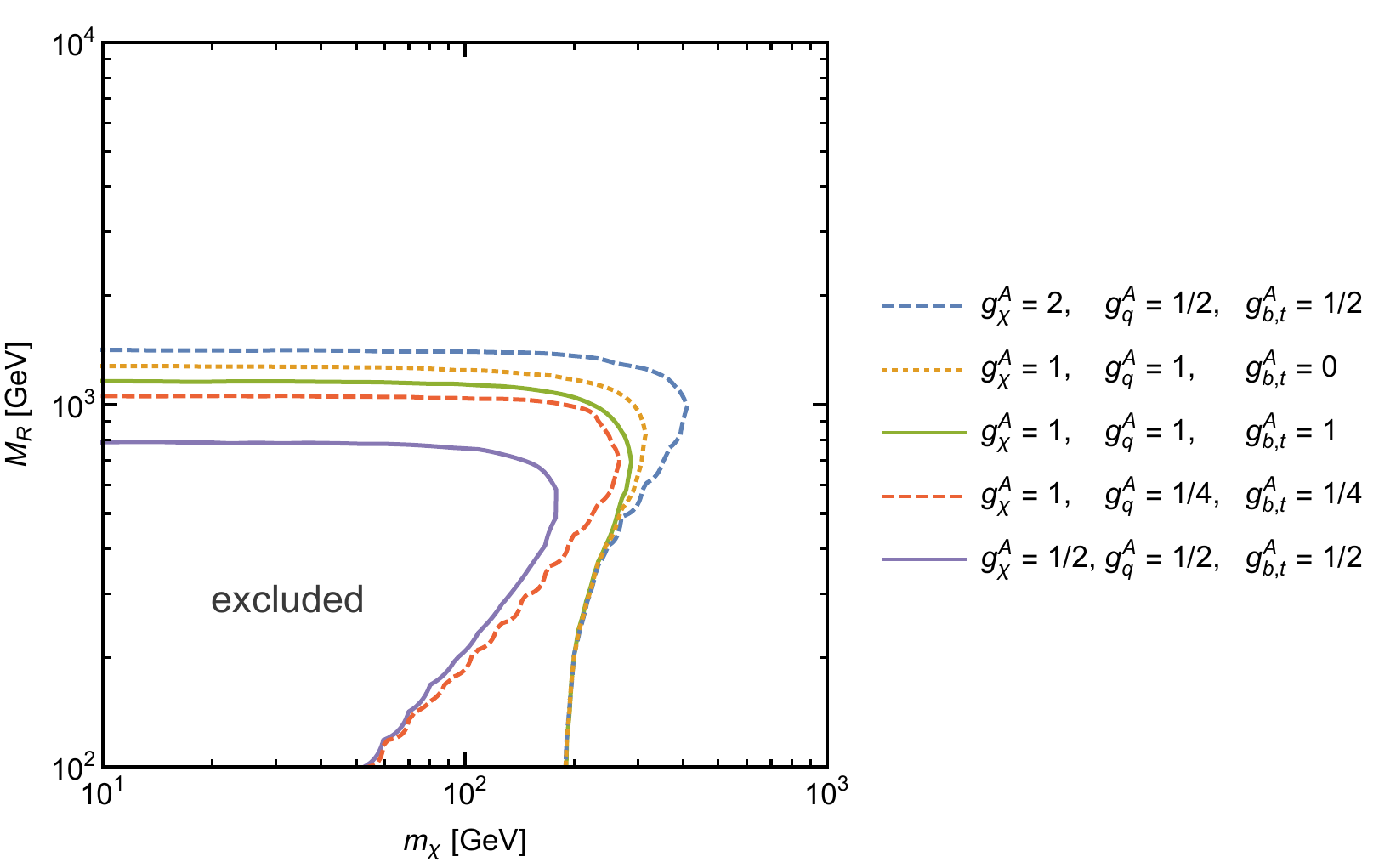}
\caption{
Exclusion bounds from the CMS monojet search in the $m_\chi$-$M_R$ parameter plane for different coupling configurations.
}
\label{fig:monojets}
\end{figure}

Across the entire parameter range that we consider, the strongest bound is obtained for the requirement $\slashed{E}_T > 450\:\text{GeV}$.\footnote{For very small values of $m_\chi$ and $M_R$ one would expect that a significantly weaker cut on $\slashed{E}_T$ gives the strongest bound, due to the missing transverse energy spectrum being rather soft. However, due to an upward fluctuation in the data at low $\slashed{E}_T$ and a downward fluctuation at high $\slashed{E}_T$ the observed bound is stronger for tighter cuts.} For this particular cut, the CMS results exclude new contributions to the fiducial monojet production cross section in excess of $7.8\:\text{fb}$ at $95\%$ C.L.

The resulting bounds for the different scenarios introduced in section~\ref{sec:scenarios} are shown in figure~\ref{fig:monojets}. We find these results to be in full agreement with the expectations discussed above. In particular, we observe that for small mediator masses, the bounds become independent of the mediator width and hence depend only on the product $g^A_q \, g^A_\chi$. For small DSP mass, the bounds depend sensitively on the mediator width (this is most obvious from comparing the green and the orange dotted line). However, changing both the width and the production cross section while keeping $g^A_\chi$ fixed leads to negligible changes in the limit of small DSP masses (as can be seen from comparing the green and the red dashed line). 

\section{Dijet searches}
\label{sec:dijet}

Searches for new resonances in dijet final states have been carried
out in several experiments. At present stringent bounds are provided
by UA2~\cite{Alitti:1993pn}, CDF~\cite{Aaltonen:2008dn},
CMS~\cite{Khachatryan:2015sja} and ATLAS~\cite{Aad:2014aqa,
  TheATLAScollaboration:2013gia} analyses, some of which have already
been applied in theoretical studies to constrain scenarios with an
additional $U(1)'$ gauge group (see~e.g.~\cite{An:2012va,
  An:2012ue,Dobrescu:2013cmh,Chiang:2015ika}). In the present section
we recast these experimental analyses in the context of an axial-vector
mediator $R$. As we will see, these constraints are complementary to
the ones described above, being more stringent for large values of
$g^A_{q}$ and small values of $g^A_{\chi}$ (or large DSP masses).

Interestingly, no single experiment provides the strongest bound
across the entire parameter space under consideration. The reason is
that for small mediator masses the QCD background, produced mainly by
two gluons in the initial state, increases much faster with increasing
centre-of-mass energy than the signal. Consequently, at the LHC the
signal is overwhelmed by QCD events in this parameter region. The
latest ATLAS and CMS dijet analyses therefore focus mostly on the
region with dijet invariant mass $m_{jj} \gtrsim
1\:\text{TeV}$, while UA2 and the Tevatron still provide competitive
bounds for smaller masses. The impressive performance of the LHC,
however, allows to produce for the first time heavy dijet resonances
in association with other SM particles, such as $Z$ and $W$ bosons,
providing a key opportunity to suppress the QCD background. Indeed, we will
show that a recent analysis of such events in the context of
technicolour~\cite{TheATLAScollaboration:2013gia} can be recast to
give strong constraints on mediator masses below $300\:\text{GeV}$.

We implement the experimental searches by means of several public
codes. To generate dijet events at parton level at leading order we use
\texttt{MadGraph~v5}~\cite{Alwall:2014hca} with the model file produced
by \texttt{Feynrules~v2}~\cite{Alloul:2013bka} and production at
leading order using the \texttt{CTEQ 6L1} PDFs~\cite{Nadolsky:2008zw}.\footnote{In order to recast the dijet analyses, both the SM background and the observed data are taken from the experimental publications, while the signal is computed at leading order. Since next-to-leading order corrections generally lead to an enhancement of the resonance production cross section~\cite{Accomando:2010fz}, our results are expected to be conservative.} Parton level
events are first passed through
\texttt{Pythia~v6}~\cite{Sjostrand:2006za} to simulate initial and
final state radiation, fragmentation and hadronisation, and
subsequently through \texttt{Delphes~v3}~\cite{deFavereau:2013fsa} for fast detector simulation.\footnote{Note that, while detector effects give a small effect in monojet searches~\cite{Aad:2015zva}, they play an important role for the dijet analysis, because they can change the shape of the resonance.} We finally employ \texttt{FastJet~v3}~\cite{Cacciari:2011ma} for jet reconstruction, and \texttt{MadAnalysis~v5}~\cite{Conte:2014zja} for cuts and data analysis. Part of the statistical analysis is carried out by means of \texttt{MCLimit}~\cite{Junk:1999kv}. Details on each analysis are described below, starting from the ones relevant for the smallest mediator masses.\\[-2mm]

\subsection{UA2 dijet analysis}

The UA2 dijet analysis~\cite{Alitti:1993pn} is based on a data sample
of 10.9$\:\text{pb}^{-1}$ of $p\bar p$ collisions at a centre-of-mass
energy of $\sqrt{s}= 630\:\text{GeV}$. The analysis essentially
requires events with two leading jets in the central region of the
detector, with $\cos{\theta} < 0.6$, and no further jets with
transverse energy $E_T>30\:\text{GeV}$. A final cut rejects events with a
dijet invariant mass, $m_{jj}$, outside the interval $(R \pm 2\sigma)
M_R$, with $R \approx 0.95$ and $\sigma \approx 0.085$ (for the precise
values of $R$ and $\sigma$, and their dependence on $M_R$,
see~\cite{Alitti:1993pn}). As shown by the UA2 collaboration, this
last cut has an efficiency $\epsilon\approx 70\%$ for dijet events
coming from a narrow-width resonance. The analysis is sensitive to the region $130\:\text{GeV} \lesssim m_{jj}\lesssim260\:\text{GeV}$.

\begin{figure}[t!]
\centering
\includegraphics[width=0.48\textwidth, clip]{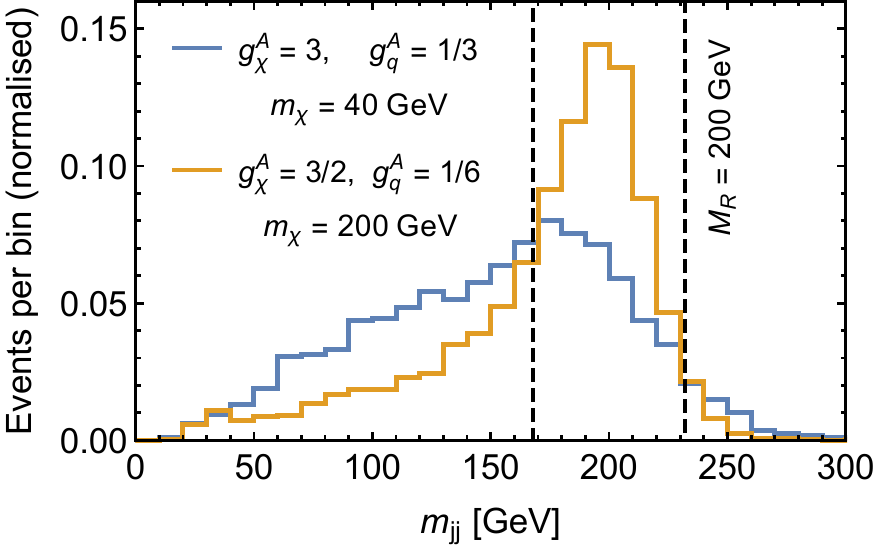}\quad
\includegraphics[width=0.48\textwidth, clip]{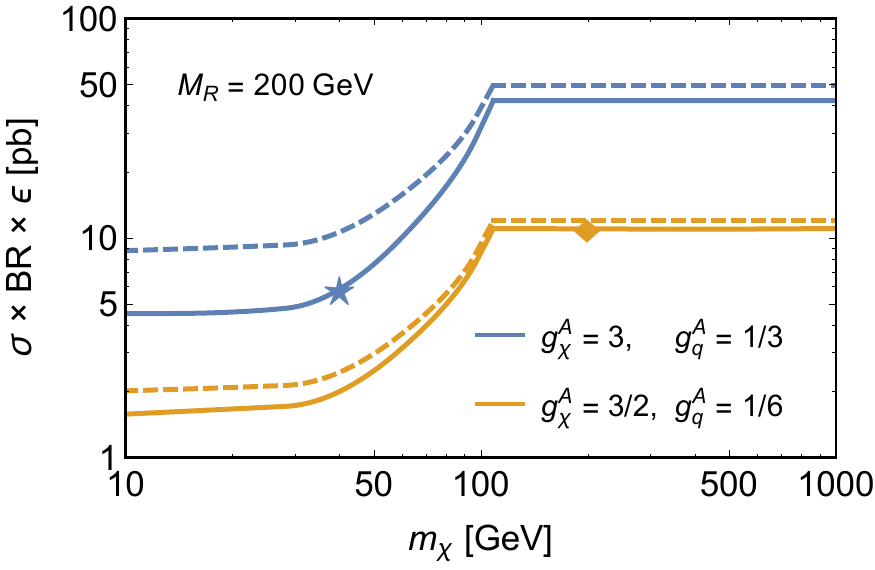}
\caption{Left: Reconstruction of the dijet invariant mass distribution for a narrow (orange) and a broad (blue) resonance with mass $M_R = 200\:\text{GeV}$ at UA2. The two vertical lines enclose the region between $(R \pm 2\sigma) m_{jj}$. Right: The solid lines indicate the dijet total production cross section times efficiency for different values of the mediator couplings with $M_R = 200\:\text{GeV}$. The parameter choices corresponding to the blue (orange) line in the left panel are represented by a blue star (orange diamond). The dashed lines are obtained by using a constant efficiency of $\epsilon = 70\%$, which is valid only for narrow resonances.}
\label{fig:eff}
\end{figure}

Implementing this analysis in our Monte Carlo simulations, we check
the efficiency $\epsilon$ and find excellent agreement with
\cite{Alitti:1993pn} whenever $\Gamma_R \ll M_R$. However, as
expected, the efficiency can be drastically smaller for broader
mediators.  This effect is displayed in the left panel of
figure~\ref{fig:eff} where we plot the (normalised) $m_{jj}$
distributions for dijet events coming from a resonance with
$M_R=200\:\text{GeV}$ and universal couplings to quarks. In the figure the
orange histogram corresponds to a narrow resonance, with $g^A_q = 1/6$,
$g^A_\chi=3/2$ and $m_\chi = M_R$ so that the invisible decay channel of
$R$ is kinematically closed. In this case the width of the mediator is
about $\Gamma_R / M_R \approx 0.01$ and hence the width of the
reconstructed distribution is dominated by final-state radiation and
detector effects, which give rise to a displacement of the peak and
tail towards the low-mass region. Despite this displacement, around
$70\%$ of the events fall within the region $(R \pm 2\sigma) M_R$
(corresponding to the interval within the vertical dashed lines) in
agreement with the UA2 finding. The blue histogram instead depicts the
scenario of a broad resonance, with $g^A_q = 1/3$, $g^A_\chi = 3$ and
$m_\chi = 40\:\text{GeV}$. In this case the larger width of $\Gamma_R
/ M_R \approx 0.1$ (convoluted with the proton and antiproton
PDFs) makes the resonance more likely to be produced at lower
invariant masses $m_{jj}$. Consequently, less than half of the signal
events fall within the $(R \pm 2\sigma) M_R$ region, and the
efficiency $\epsilon$ decreases to about $40\%$.

The dependence of $\epsilon$ on the mediator width is highlighted in figure~\ref{fig:eff} (right panel) where we plot the
total production cross section times efficiency as a function of
$m_\chi$. Curves corresponding to the parameter settings $g^A_q = 1/3$ and $g^A_\chi = 3$
($g^A_q = 1/6$ and $g^A_\chi = 3/2$) are marked in blue (orange).  Dashed
lines are obtained by assuming the (nominal) efficiency $\epsilon$
provided in~\cite{Alitti:1993pn}, whereas solid lines are calculated
using the (actual) efficiency that takes into account the broad-width
effect on the dijet invariant-mass cut. While for the narrow-width
scenario (cf.~orange curves) the discrepancy between nominal and
actual efficiency is negligible, for the broad scenario (cf.~blue
curves) the discrepancy is relevant. For instance, for the broad case
considered in the figure the ratio between nominal and actual
efficiencies, $r_\epsilon$, increases from around 1.2 to 2 as the invisible decay channel opens up.

Once $r_\epsilon$ is determined, it is straightforward to recast the
UA2 bound in the model we are analysing. The UA2 collaboration
explicitly applies its constraint to a sequential Standard Model (SSM)
$Z'$ and presents a 90\% C.L.~bound on ${\sigma(p\bar{p}\rightarrow
  Z_\text{SSM})\times\text{BR}(Z_\text{SSM}\rightarrow j j)}$
(assuming the efficiency $\epsilon\approx
0.7$)~\cite{Alitti:1993pn}. A parameter point in our model is excluded if the predicted total cross section $\sigma(p\bar{p}\rightarrow R\rightarrow j j)$ violates the condition
\begin{equation}
\frac{ r_\epsilon\, \sigma(p\bar{p}\rightarrow R\rightarrow j j)\,}
{\sigma(p\bar{p}\rightarrow Z_\text{SSM})\times\text{BR}(Z_\text{SSM}\rightarrow j j)} < 1~.
\end{equation}
The region of the $M_{R}$-$m_\chi$ plane excluded by this bound is
shown in figure~\ref{fig:dijets} (blue regions with solid borders),
where different coupling scenarios are assumed in the various panels.\footnote{We find the bounds from UA2 to be somewhat stronger than the ones shown in~\cite{Dobrescu:2013cmh}, in agreement with the analyses presented in~\cite{An:2012ue,Chiang:2015ika}.}
For large couplings $g^A_q$ the UA2 constraint rules out the mass range $130\:\text{GeV} < M_R < 260\:\text{GeV}$. 
For $g^A_q\approx 1/4$ the weakening in sensitivity of the UA2 experiment
at $m_{jj}\sim 200\:\text{GeV}$ leads to an allowed horizontal band. 
The UA2 constraint is not sensitive to a
resonance with $g^A_q\lesssim 1/6$ for any value of $m_\chi$. Moreover, for sufficiently large $g^A_\chi$, the UA2 bound
cannot constrain the region with $m_\chi\lesssim
M_R/2$.

\subsection{ATLAS lepton-associated dijet analysis}

The ATLAS collaboration has analysed a 20.3\,fb$^{-1}$ data sample
searching for dijet resonances produced in association with a gauge
boson via $p p \rightarrow R\, Z\rightarrow j j \ell^+ \ell^-$ and $p
p \rightarrow R\, W^\pm\rightarrow j j \ell^\pm\nu$ processes at
$\sqrt{s}=8\:\text{TeV}$~\cite{TheATLAScollaboration:2013gia}. The
presence of at least one lepton in the final state allows to highly
reduce the QCD background and therefore to probe the region with dijet
invariant mass $130 \:\text{GeV}< m_{jj} < 300\:\text{GeV}$ which is
otherwise inaccessible at the LHC. We implement these two ATLAS
analyses to complement the UA2 constraint in the low $M_R$ region.

We produce $p p \rightarrow R\, Z\rightarrow j j \ell^+ \ell^-$ and $p p \rightarrow R\, W^\pm\rightarrow j j \ell^\pm\nu$ Monte
Carlo events in accordance with ATLAS detector specifications. We also
impose the cuts enlisted in the two analyses. The search for the
  $Z$-($W$-)associated production requires
  $p_T^{\ell\ell(\ell\nu)}>50\:\text{GeV}$ where
  $p_T^{\ell\ell(\ell\nu)}$ is the transverse momentum of the dilepton (lepton
  and $\slashed{E}_T$) system. In addition, both analyses
  require two jets with $p_T>30\:\text{GeV}$ and relative
pseudorapidity $|\Delta\eta_{jj}|<1.75$
(see~\cite{TheATLAScollaboration:2013gia} for details on the cuts). To
reconstruct the jets we employ an anti-$k_t$~\cite{Cacciari:2008gp}
algorithm with $R = 0.4$. As in the UA2 analysis, the final outcome of
the event analysis is a distribution of dijet invariant masses across
several $m_{jj}$ bins.

To validate the simulations we produce the dijet
invariant-mass distribution for the corresponding SM background
(mainly $Z$+jets and $W$+jets production) and we find
agreement with the one reported
in~\cite{TheATLAScollaboration:2013gia}. As a last step, we compare
the $m_{jj}$ distributions observed by
ATLAS with those predicted by the SM with and without a mediator dijet
signal.\footnote{The numbers of observed events in each bin have been
  digitised from the plot in~\cite{TheATLAScollaboration:2013gia} in
  order to properly estimate the bounds on our theory. Indeed, the
  bound cannot be directly read off from the exclusion limits provided
  in \cite{TheATLAScollaboration:2013gia} because the analysis is
  devoted to a different model.}  We quantify the (in)compatibility of
prediction and observation by means of a CL$_s$ statistical analysis
considering ATLAS measurements on $W$- and $Z$-associated
  productions at once.\footnote{For the basics of the implementation of the
CL$_s$ method, see e.g.\ appendix C of~\cite{delAguila:2013mia}. For the combination procedure we
follow~\cite{deBlas:2013qqa}.} More
precisely we determine the confidence level that the signal events
distributed in all $m_{jj}$ bins are incompatible with the observed
distribution.  This statistical approach enables us to properly
analyse even broad resonances.

\begin{figure}[t!]
\centering
\includegraphics[width=0.42\textwidth, clip]{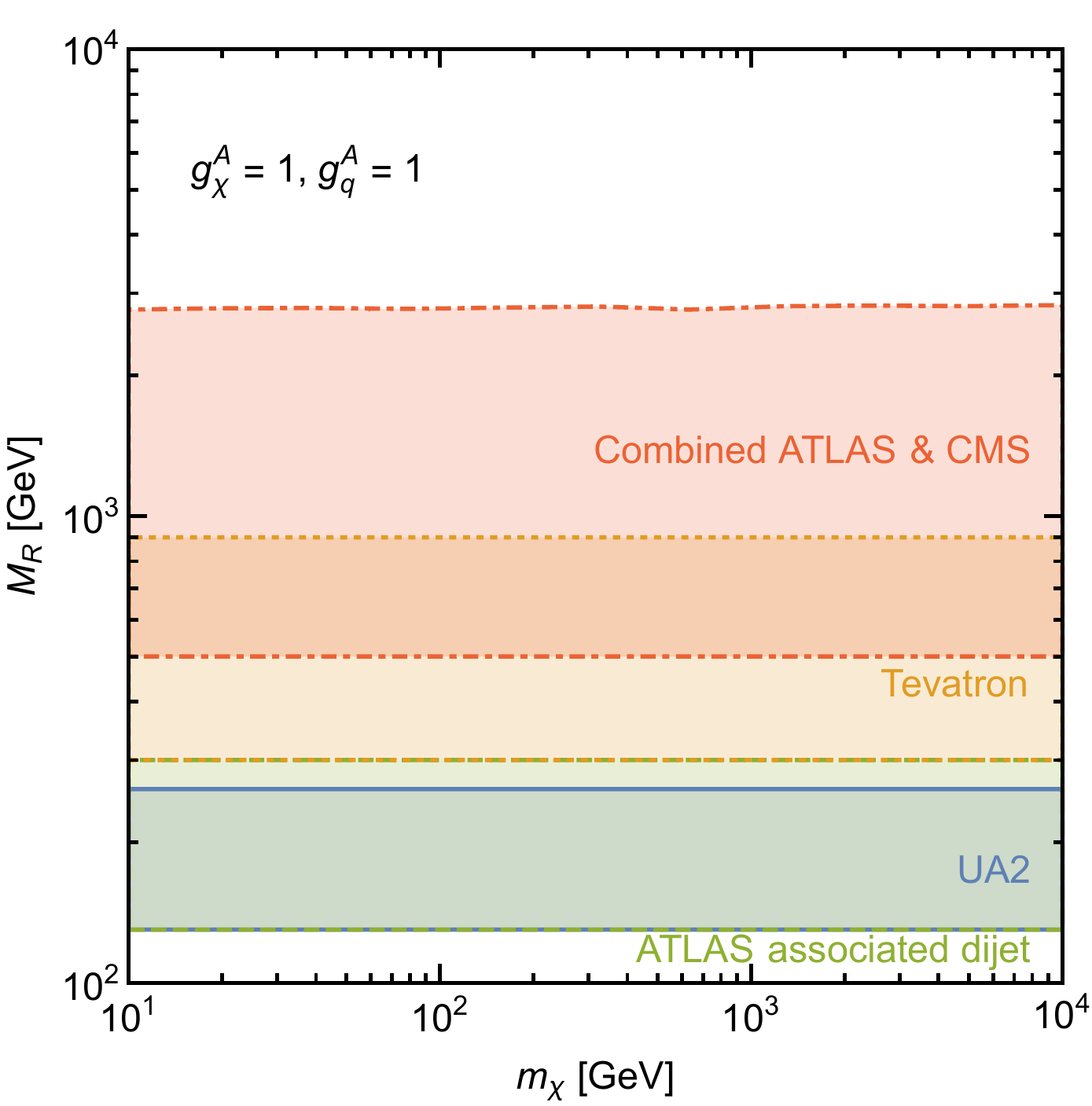}\quad
\includegraphics[width=0.42\textwidth, clip]{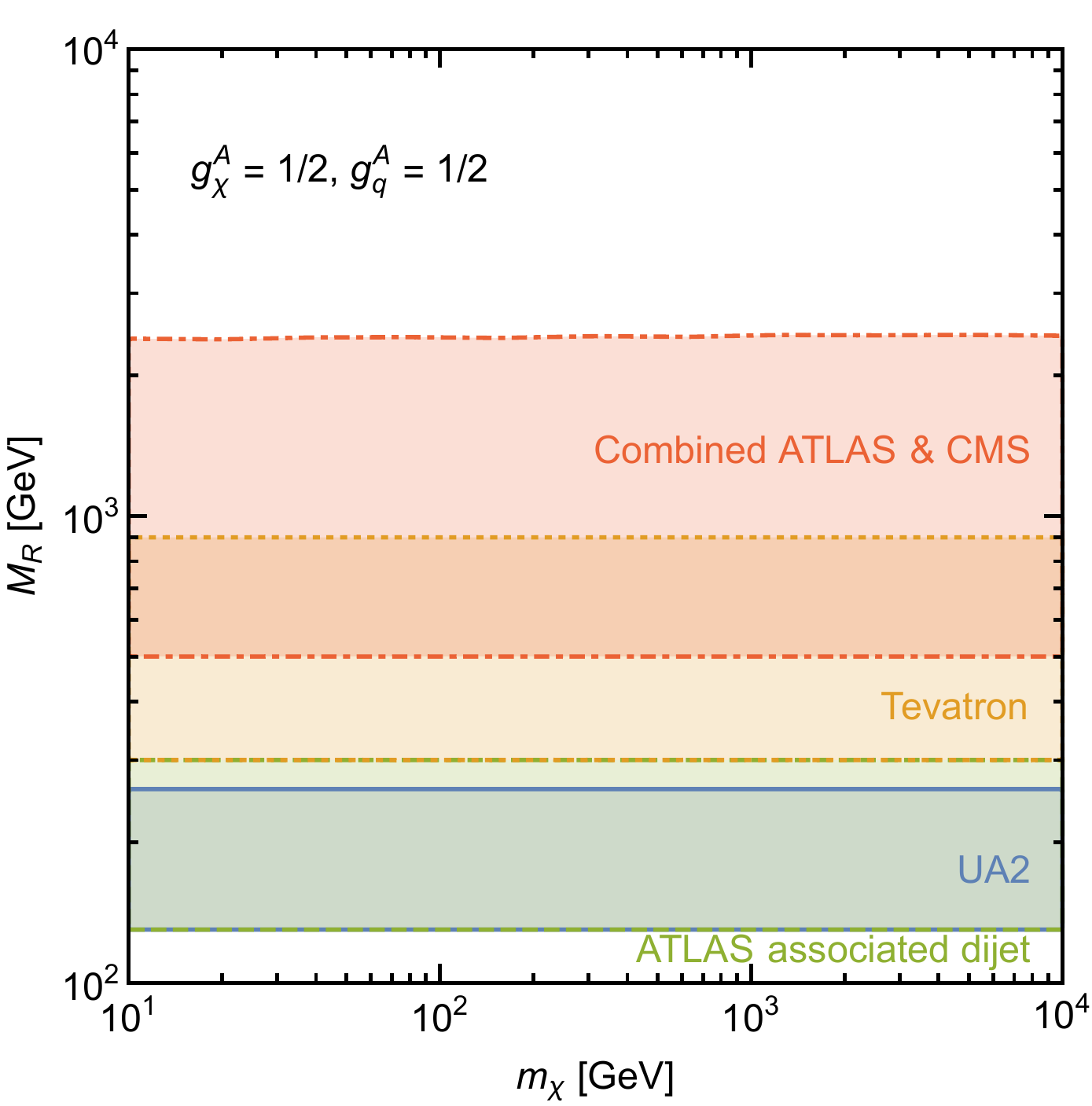}

\includegraphics[width=0.42\textwidth, clip]{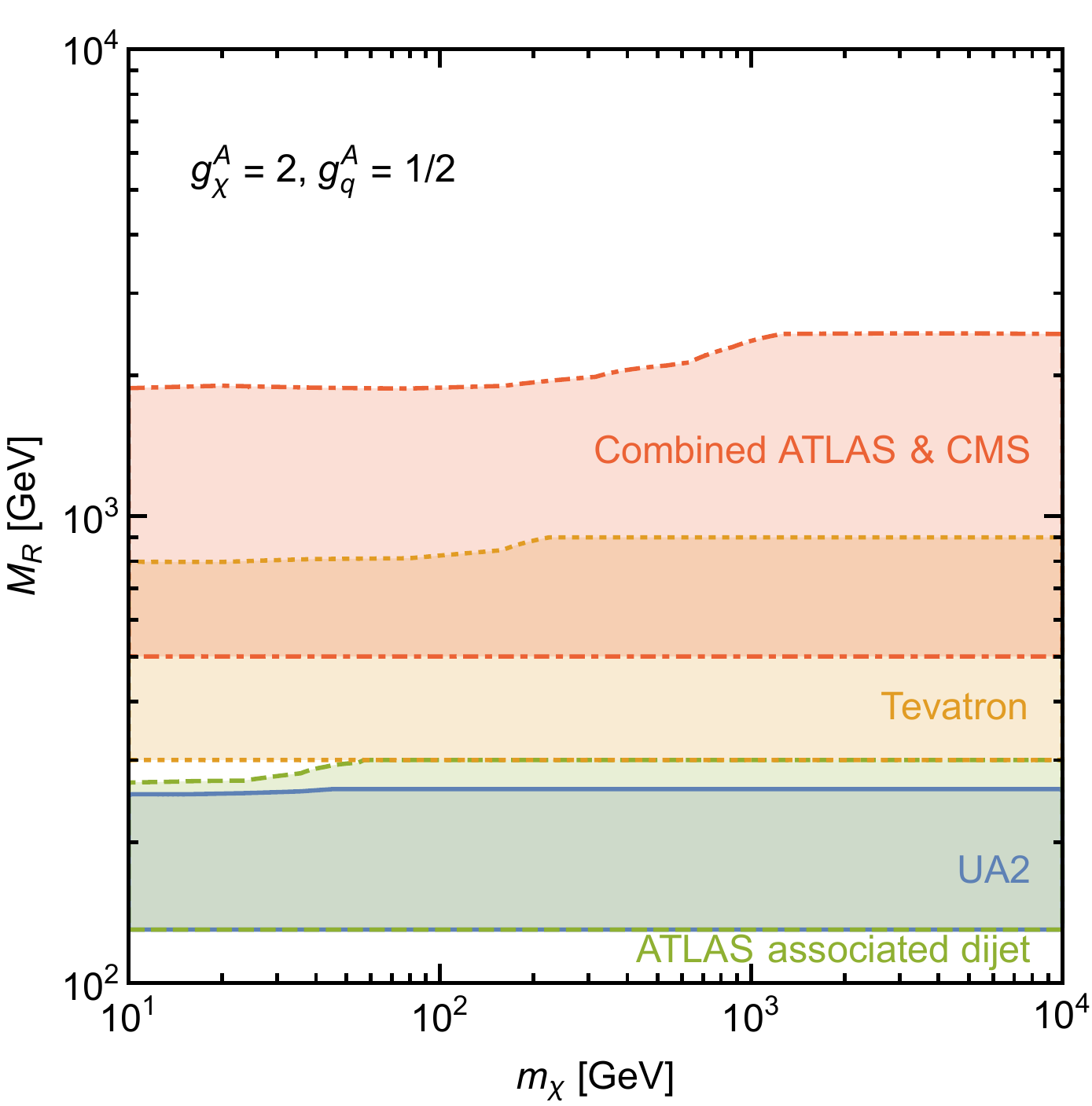}\quad
\includegraphics[width=0.42\textwidth, clip]{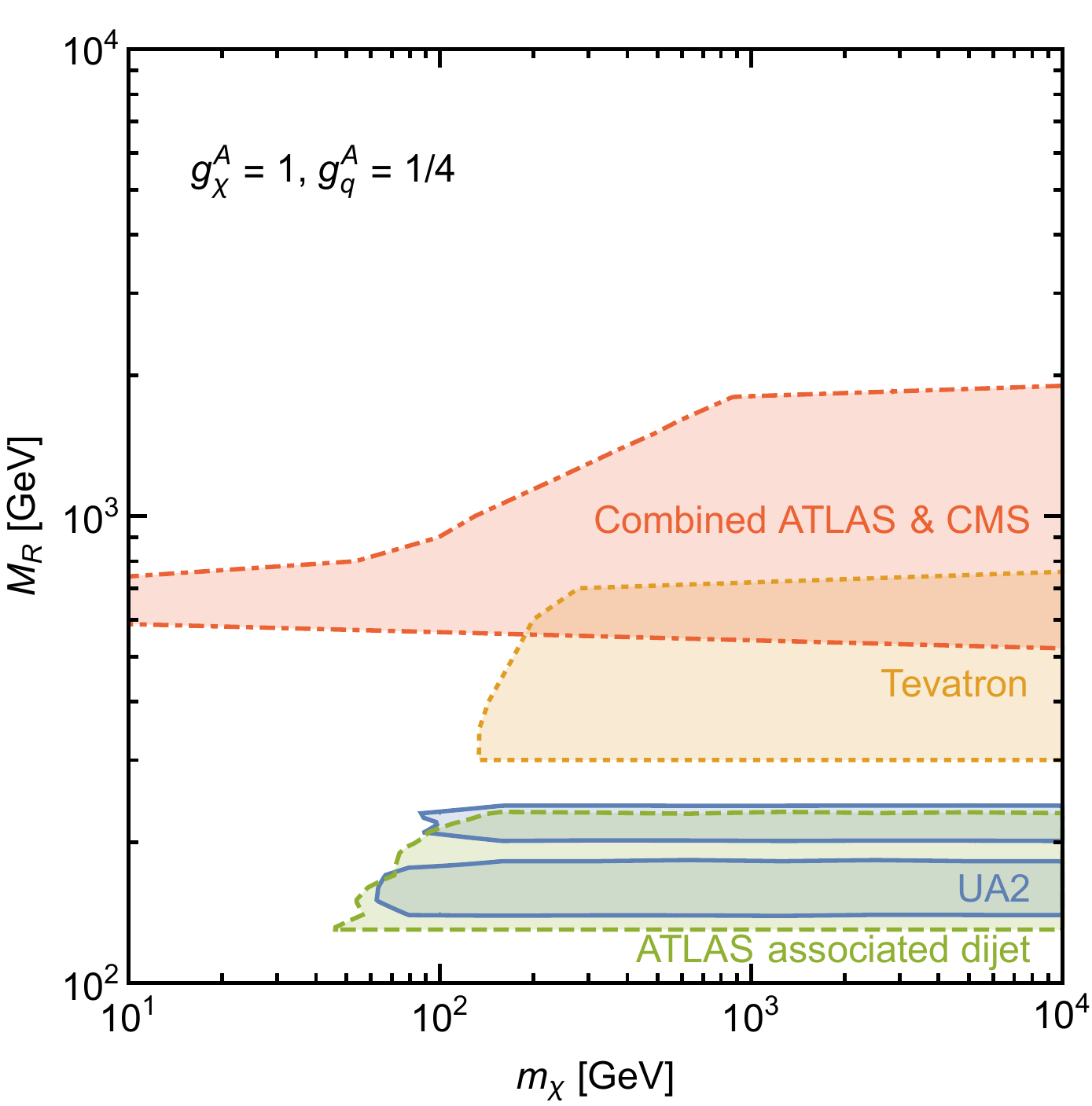}

\includegraphics[width=0.42\textwidth, clip]{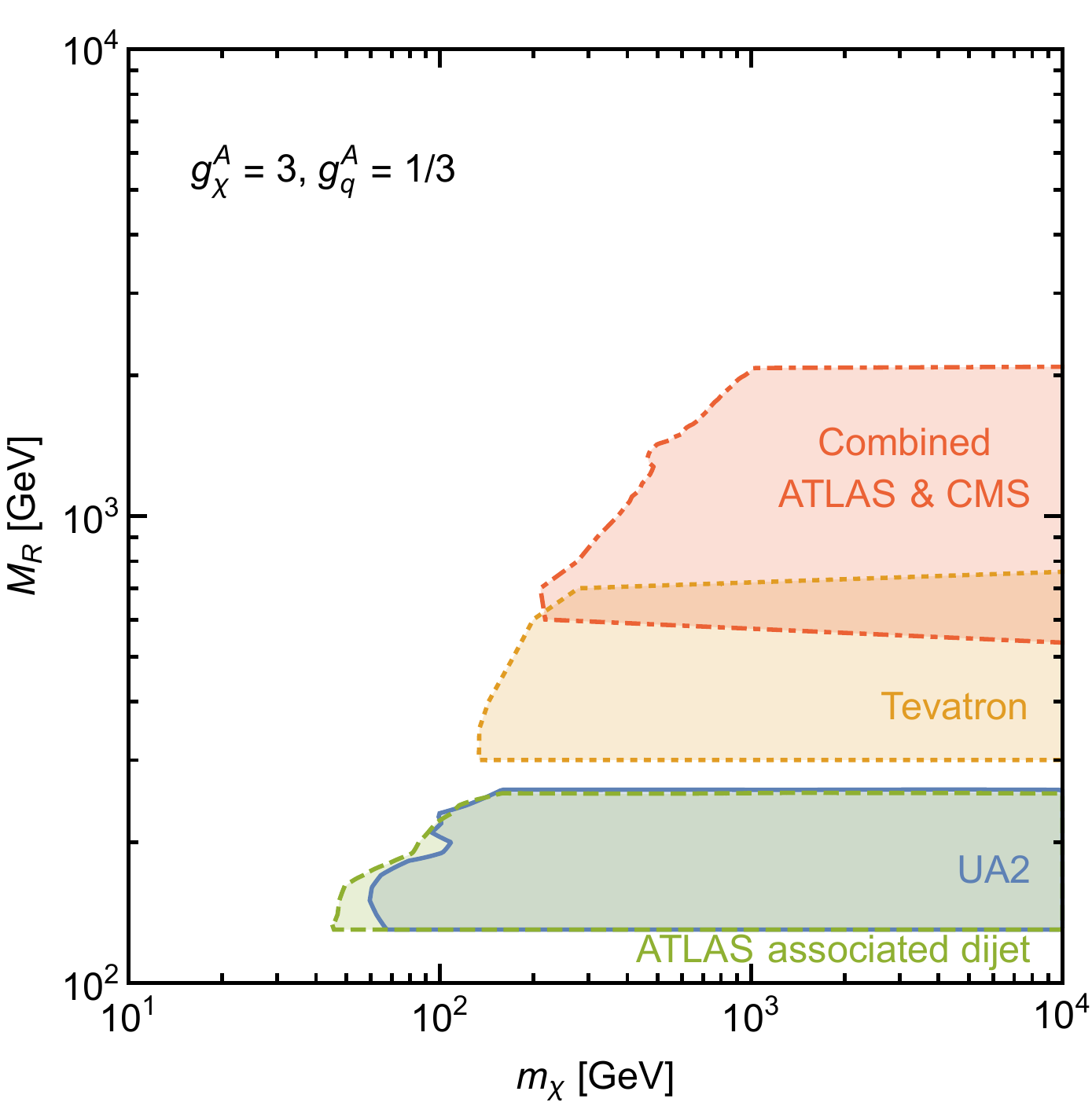}\quad
\includegraphics[width=0.42\textwidth, clip]{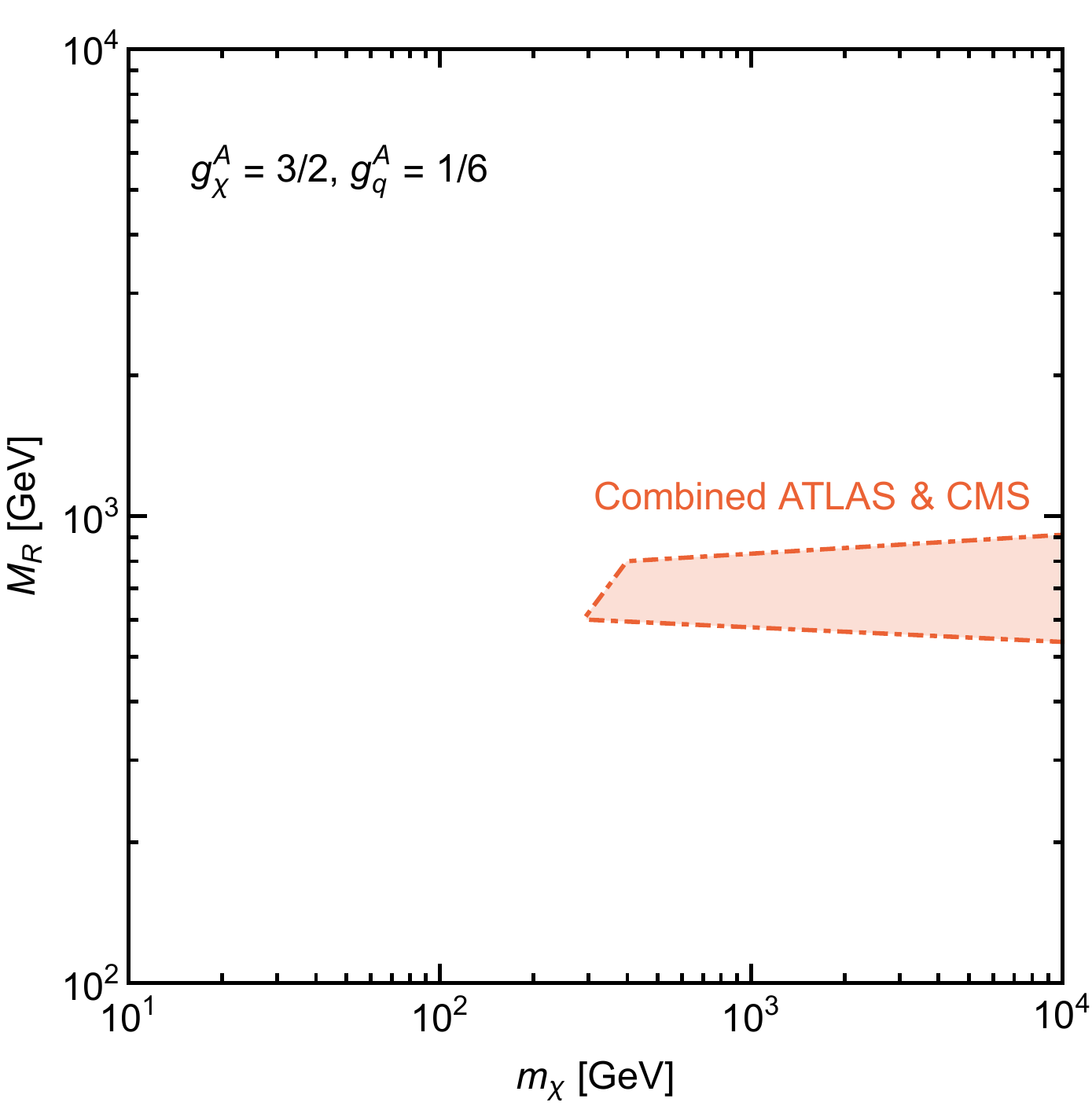}
\caption{Parameter regions excluded by dijet searches at UA2, the Tevatron and the LHC. For the left (right) column, we have fixed $g \equiv (g^A_\chi \, g^A_q)^{1/2} = 1$ (g = 0.5), while the different rows show different coupling ratios $g^A_\chi / g^A_q$. The UA2 bounds correspond to 90\% C.L., all other bounds are shown at 95\% C.L.}
\label{fig:dijets}
\end{figure}

The final result is presented in figure~\ref{fig:dijets} (green
regions with dashed borders). We see that the ATLAS 95\% C.L.~bound
turns out to be slightly stronger than the one by UA2. Notably,
the region at $M_R\simeq 200\:\text{GeV}$ where UA2 loses sensitivity for $g^A_q=1/4$
is now clearly ruled out.

\subsection{CDF dijet analysis}

The CDF collaboration performed searches for a dijet resonance in $p \bar p$
collision data with an integrated luminosity of $1.13\:\text{fb}^{-1}$
at $\sqrt{s}=1.96\:\text{TeV}$~\cite{Aaltonen:2008dn}. This analysis
requires two jets within the central region of the detector,
corresponding to a rapidity $|y|<1$. These jets are then reconstructed
using a cone jet-clustering algorithm with $R = 0.7$. While the search in principle constrains dijet resonances in the 
mass range $260\:\text{GeV} < m_{jj} <
1400\:\text{GeV}$, we focus on the region $300\:\text{GeV} \le m_{jj} \le 900 {\:\text{GeV}}$, where the most stringent
constraints are expected~\cite{Dobrescu:2013cmh}. 
The lower end of this mass region is the most interesting, because
the mediator is already too heavy to be produced in sufficient
number for the UA2 and ATLAS searches above, but is still too light to be disentangled from the SM background in LHC dijet searches.

For our analysis, we mimic the CDF detector characteristics and recast
the cuts employed in the experimental analysis. We validate the
analysis by reproducing the SM background distribution and checking
for agreement with the background reported
in~\cite{Aaltonen:2008dn}. To determine the compatibility of the model
with the CDF data, we apply an analysis analogous to the
one discussed above for the lepton-associated dijet
search. Given the large amount of events at small invariant masses and the various error sources affecting the correct description of the QCD background, we also include systematic uncertainties in the normalisation of the background in our statistical analysis. The systematic uncertainties have been (globally) fixed in such a way that the CDF bounds can be reproduced when considering a narrow resonance.

The results of our simulations are presented in
figure~\ref{fig:dijets}. The CDF 95\% C.L.~bounds (orange regions with
dotted borders) strongly constrain the $M_R$-$m_\chi$ parameter
space. The general trend is that for $0.25\lesssim g^A_q \lesssim1$ and
$300\:\text{GeV}\lesssim M_R\lesssim 700\:\text{GeV}$ only the parameter space with a large
mediator invisible width evades the CDF bound because of a twofold effect:
for $m_\chi<M_R/2$ and large enough $g^A_\chi$ the mediator branching
ratio to quarks is reduced and its total width broadens, resulting in
a loss of sensitivity.

\subsection{ATLAS and CMS dijet analyses}

The CMS and ATLAS collaborations look for dijet resonances in
19.7\,fb$^{-1}$ and 20.3\,fb$^{-1}$ of data of $pp$ collisions at
$\sqrt{s}=8\,$TeV,
respectively~\cite{Khachatryan:2015sja,Aad:2014aqa}. The CMS analysis
is somewhat more involved than the ones discussed above. It first
reconstructs jets by means of an anti-$k_T$ algorithm with $R<0.5$ and
selects jets with $p_T>30\:\text{GeV}$ and $\eta<2.5$. Second,
geometrically close jets ($\Delta R<1.1$) are combined into wide jets,
which are then selected to form a final (wide) dijet system fulfilling
$|\Delta \eta_{jj}|<1.5$, $m_{jj}>890\:\text{GeV}$ and $H_T\equiv\sum_i
p_T^{j_i} > 650\:\text{GeV}$. The ATLAS analysis is comparably straightforward. It
uses the anti-$k_T$ algorithm with $R<0.5$ to reconstruct jets and
rejects those with $p_T<50\:\text{GeV}$. Jets are then joined into a
dijet system whose leading jets must have rapidity $|y|<2.8$ and $|y^*|\equiv |y_\text{lead}-y_\text{sublead}|/2 <0.6$. Finally, only dijets with an invariant mass of
$m_{jj}>250\:\text{GeV}$ are considered.

We analyse the ATLAS and CMS searches for mediator masses
$M_{R}>500\:\text{GeV}$ (for smaller masses the LHC loses sensitivity
compared to the Tevatron). To validate the simulation and data
analysis we essentially proceed as explained for the CDF and ATLAS
lepton-associated searches above. In particular we employ the
\texttt{MCLimit} code for both ATLAS and CMS results at once, assuming
that they are not correlated. We find that the ATLAS and CMS 95\%
C.L.~exclusion limits are very powerful (cf.~red regions inside
dot-dashed borders in figure~\ref{fig:dijets}) although they can be
avoided in the parameter region with large invisible width.  
We also observe from figure~\ref{fig:dijets} that even in the absence of invisible decays (i.e.\ for $M_R < 2 \, m_\chi$) the bounds on $M_R$ become stronger as $g_q^A$ is increased. The reason is that the enhancement of the $R$ production cross section overcompensates the reduction of the detection efficiency due to the broadening of the resonance.

\begin{figure}[t!]
\centering
\includegraphics[width=0.46\textwidth, clip]{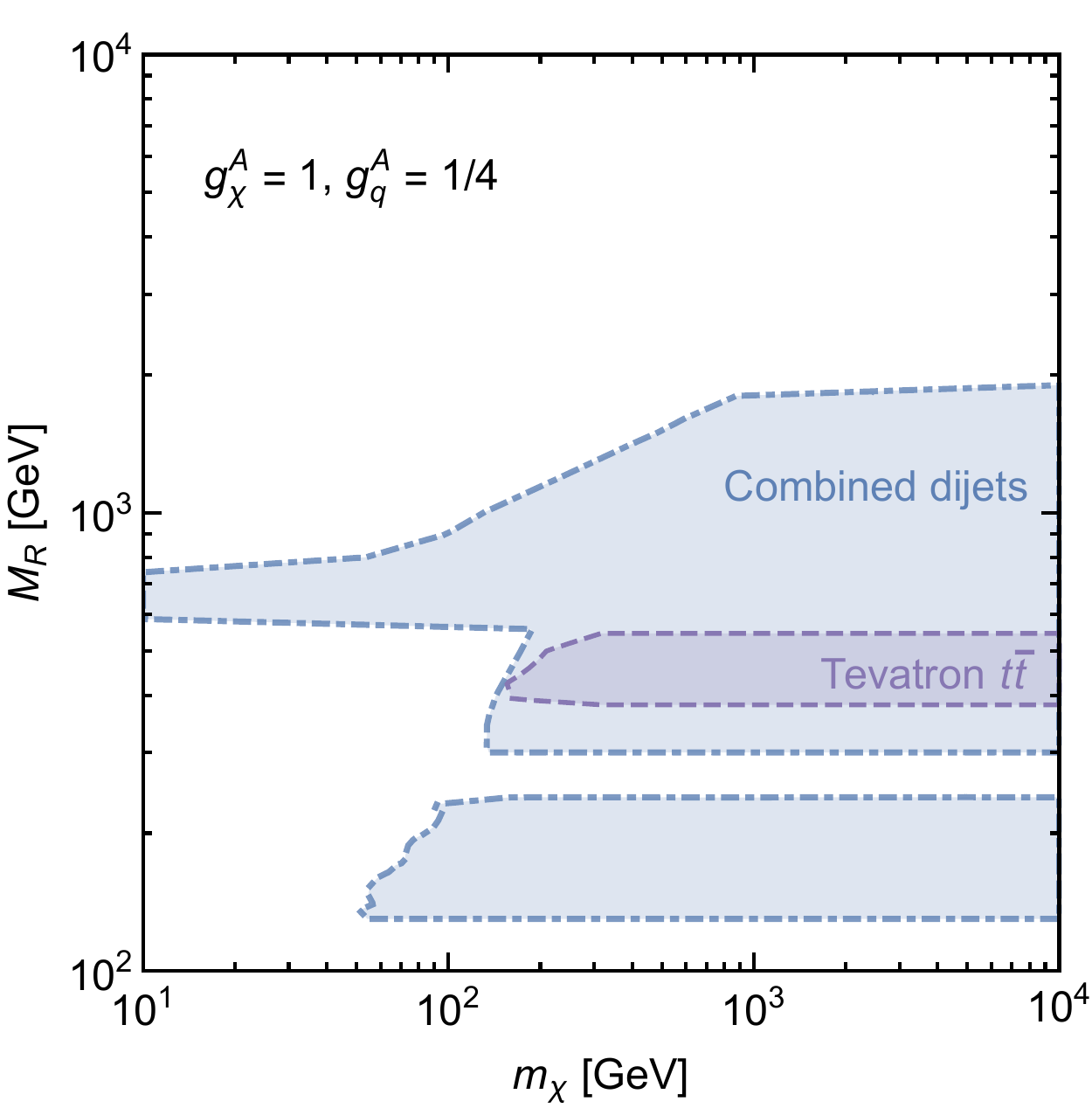}\quad 
\includegraphics[width=0.46\textwidth, clip]{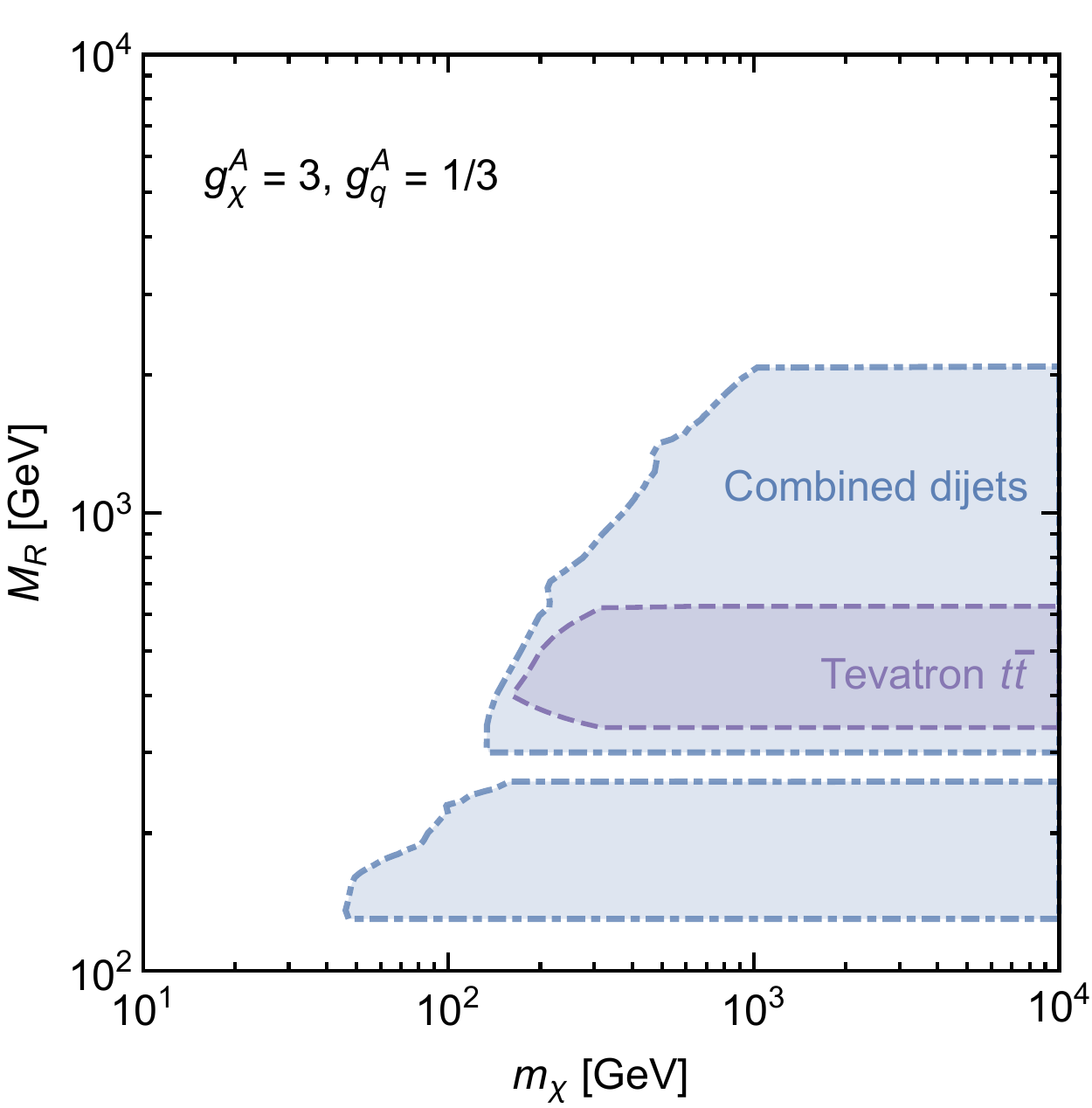}
\caption{Comparison of the combined bounds from searches for dijet resonances (blue) and the constraints from the Tevatron measurements of the total $t\bar{t}$ production cross section (purple) for two particular coupling combinations.}
\label{fig:ttbar4}
\end{figure}

\subsection{Comparison with top-quark pair production at the Tevatron}

Given that searches for dijet resonances yield relatively weak
constraints for $300\:\text{GeV} < M_R < 500\:\text{GeV}$, it is
interesting to consider other kinds of searches that can potentially
probe this particular parameter region. Indeed, if the vector mediator
has non-zero couplings to the third generation of quarks, its presence
can lead to a significant enhancement of the total $t\bar{t}$
production cross section. This cross section has been precisely
determined by combining CDF and D0 measurements at
$\sqrt{s}=1.96\:$TeV with $\mathcal L=8.8\:$fb$^{-1}$~\cite{ttbar},
leading to constraints that are most relevant if $M_R$ is close to the
$t\bar{t}$ threshold.\footnote{Note that recent LHC measurements of
  the total $t \bar{t}$ production cross section, such
  as~\cite{Khachatryan:2014loa}, do not improve on the constraints
  from the Tevatron, since $t \bar{t}$ production at the LHC is dominated
  by gluons.}

To obtain the constraints from the Tevatron we simulate $t\bar{t}$
production at leading order, taking into account the interference
between the SM contribution and new physics effects. If the
contribution of the new vector mediator is negligible, we obtain
$\sigma(p\bar{p} \rightarrow t\bar{t}) = 6.03\:\text{pb}$. Comparing
this value to recent NNLO+NNLL calculations of $t\bar{t}$ production
at the Tevatron, which quote $\sigma(p\bar{p} \rightarrow t\bar{t}) =
7.35^{+0.28}_{-0.33}\:\text{pb}$~\cite{Czakon:2013goa}, allows us to
determine the relevant $K$-factor to be $K = 1.22 \pm 0.05$. We can
then approximately capture the effect of higher-order corrections by
rescaling all simulated cross sections by this factor. 

The combined results of several $t\bar{t}$ measurements at the Tevatron
determine the $t\bar{t}$ production cross section to be
$\sigma(p\bar{p} \rightarrow t\bar{t}) =
7.60\pm0.41\:\text{pb}$~\cite{ttbar}. Combining experimental and
theoretical uncertainties, we can therefore place an upper bound on
the rescaled cross section of $\sigma(p\bar{p} \rightarrow t\bar{t}) <
8.6\:\text{pb}$ at 95\% C.L.

As expected, the resulting constraints are most relevant for mediator
masses just above the $t\bar{t}$ threshold, i.e. $M_R \sim
400\:\text{GeV}$. Moreover, just like the constraints from dijet
resonance searches, we find the strongest constraints if either
$m_\chi > M_R / 2$ or $g_q^A \gtrsim g^A_\chi$ such that the mediator
decays dominantly into quarks. In other words, measurements of the
$t\bar{t}$ cross section constrain the same parameter space also
probed by searches for dijet resonances at the Tevatron and we find that
the latter constraint is always slightly stronger. Two particular
examples are shown in figure~\ref{fig:ttbar4}, where we compare the
combined dijet bounds from figure~\ref{fig:dijets} (blue) with the
parameter region excluded by $t\bar{t}$ measurements (purple). For
$g_q^A\lesssim 1/6$ we do not obtain any constraints from $t\bar{t}$
measurements. Since the constraints from $t\bar{t}$ do not improve
upon the constraints discussed above, we will not show them in the
remaining plots.

Nevertheless, it is worth emphasising that bounds on the total
$t\bar{t}$ cross section do provide an important independent constraint
for large quark couplings. We find that for \mbox{$g^A_q = 0.5$} (and
either small $g^A_\chi$ or large $m_\chi$) the Tevatron can
essentially exclude the mass range $330\:\text{GeV} \lesssim M_R
\lesssim 700\:\text{GeV}$, while for $g^A_q = 1$ the entire range
$ M_R \lesssim 900\:\text{GeV}$ is
excluded. In the latter case, the $t\bar{t}$ cross section can be as
large as four times the SM prediction (for $M_R \sim
400\:\text{GeV}$), making it clear that this case is solidly ruled
out.

\section{Combined constraints}
\label{sec:combination}

\begin{figure}[t!]
\centering
\includegraphics[width=0.41\textwidth, clip]{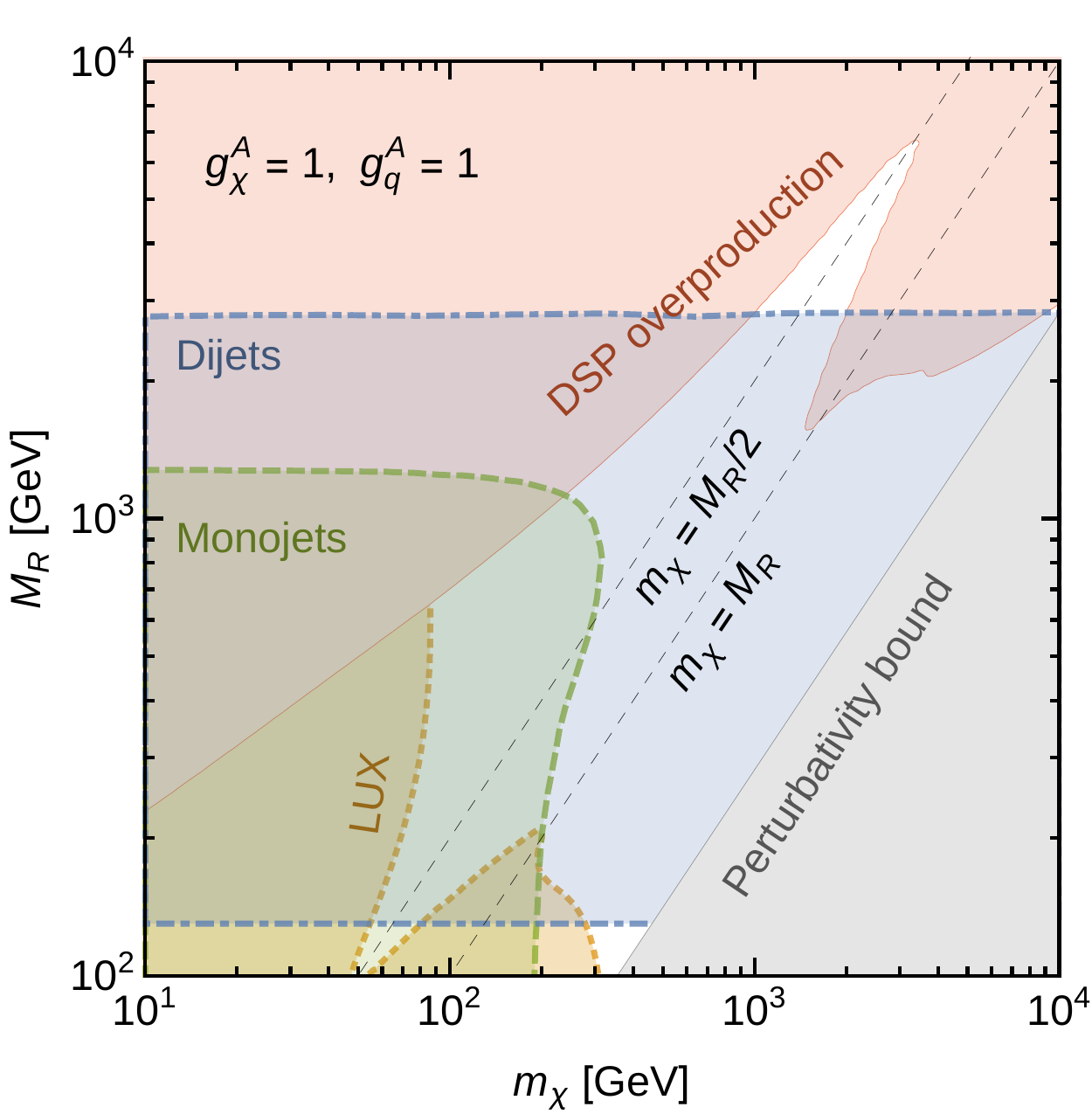}\quad
\includegraphics[width=0.41\textwidth, clip]{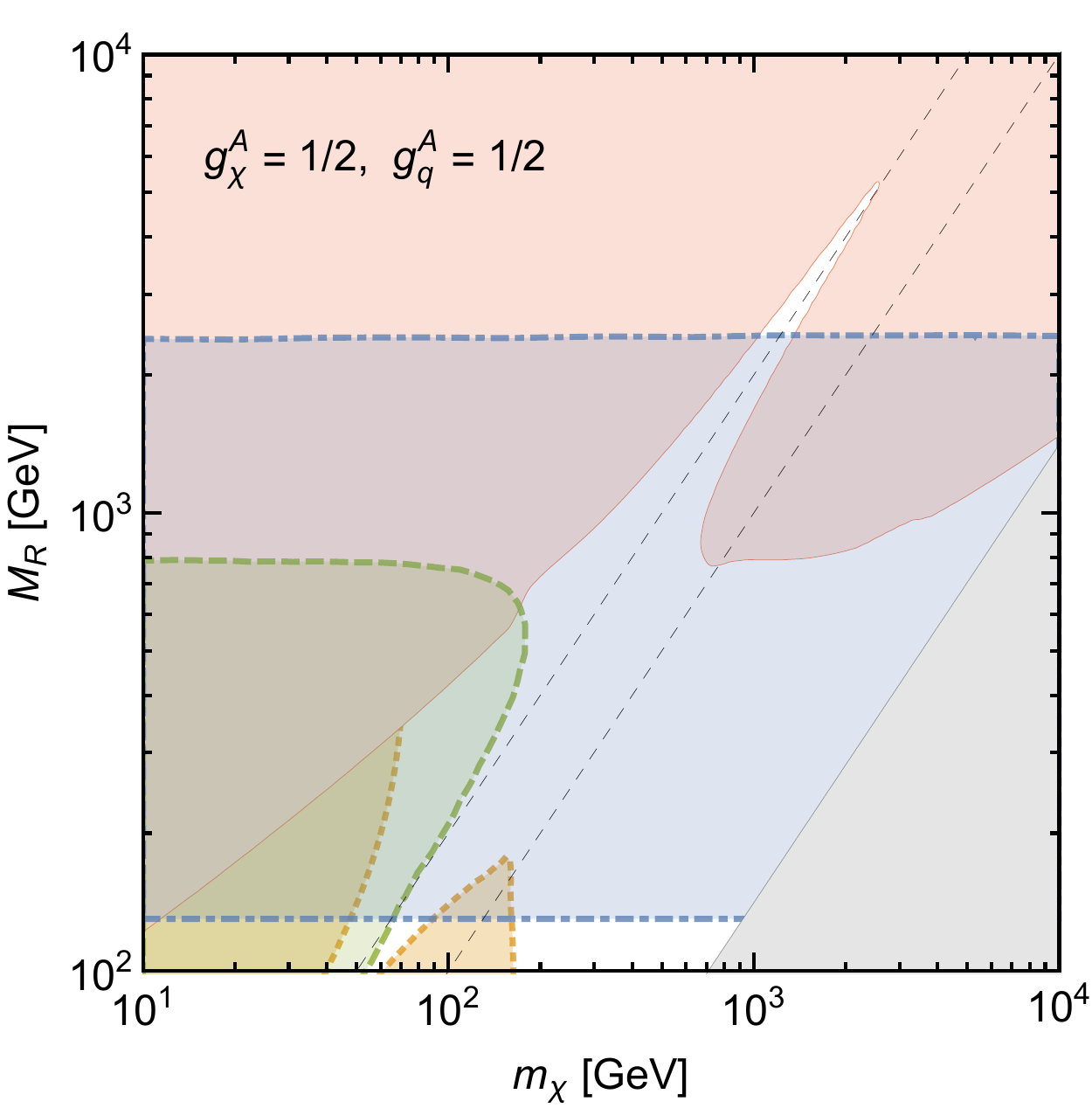}

\includegraphics[width=0.41\textwidth, clip]{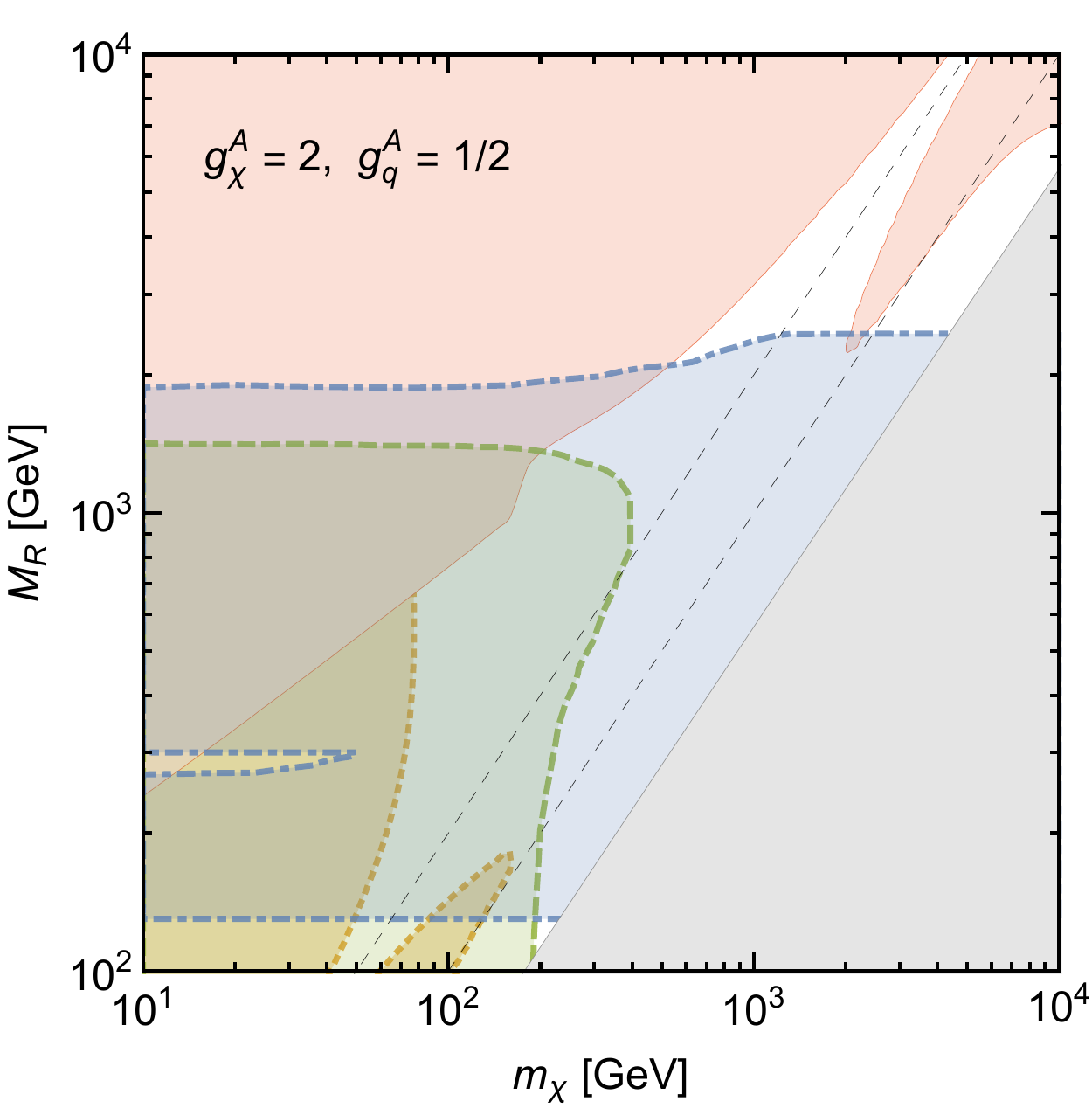}\quad
\includegraphics[width=0.41\textwidth, clip]{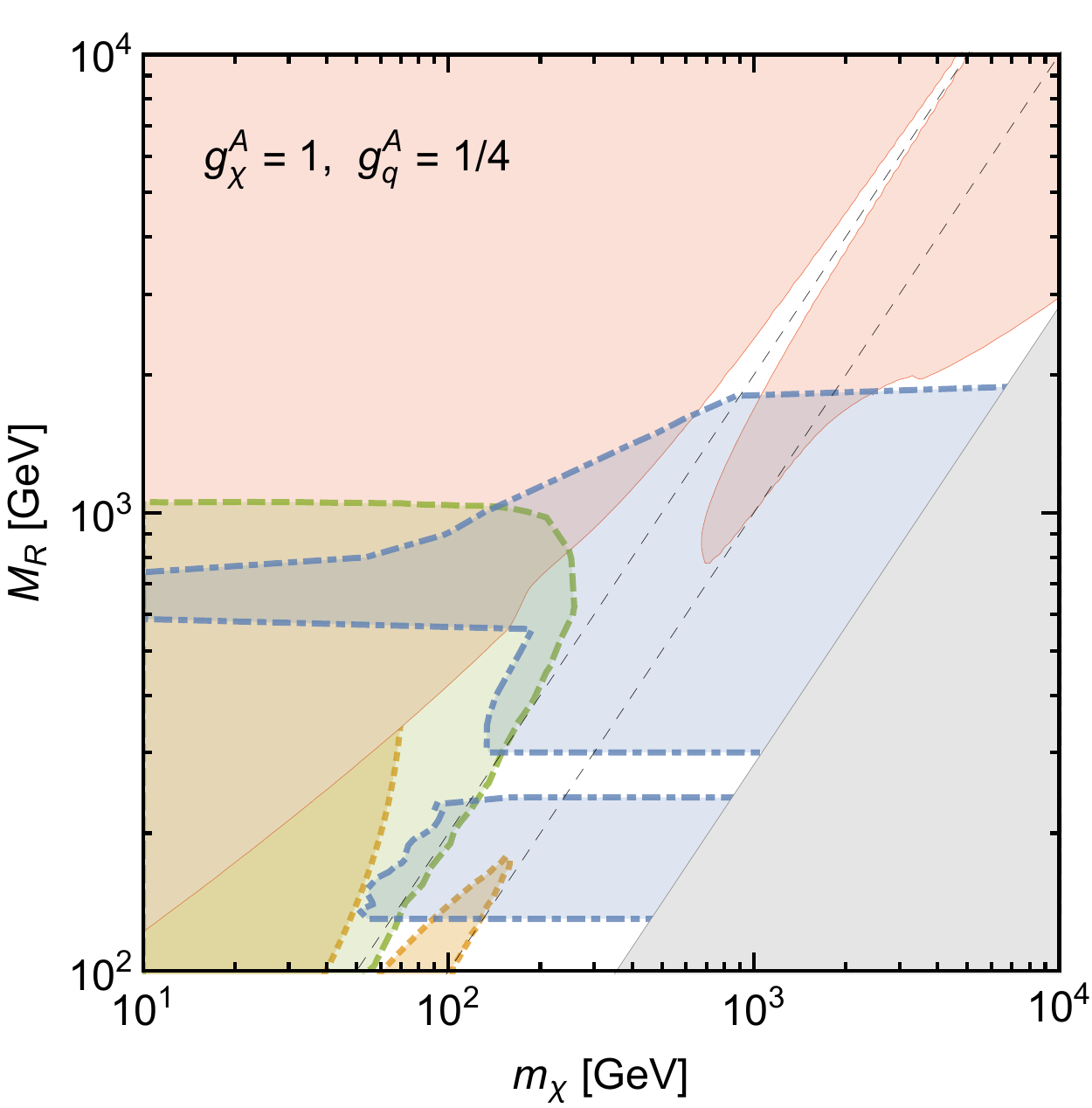}

\includegraphics[width=0.41\textwidth, clip]{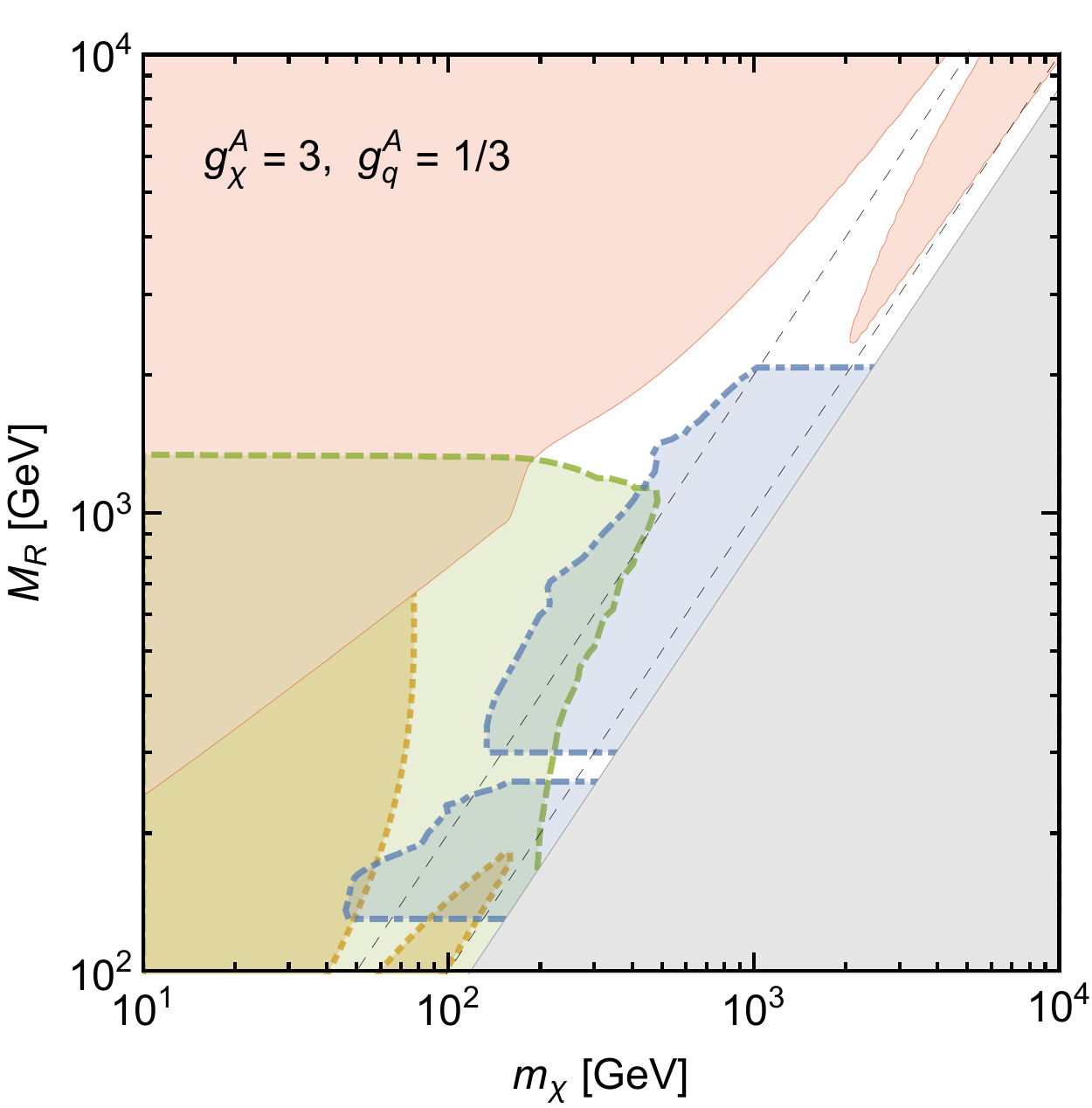}\quad
\includegraphics[width=0.41\textwidth, clip]{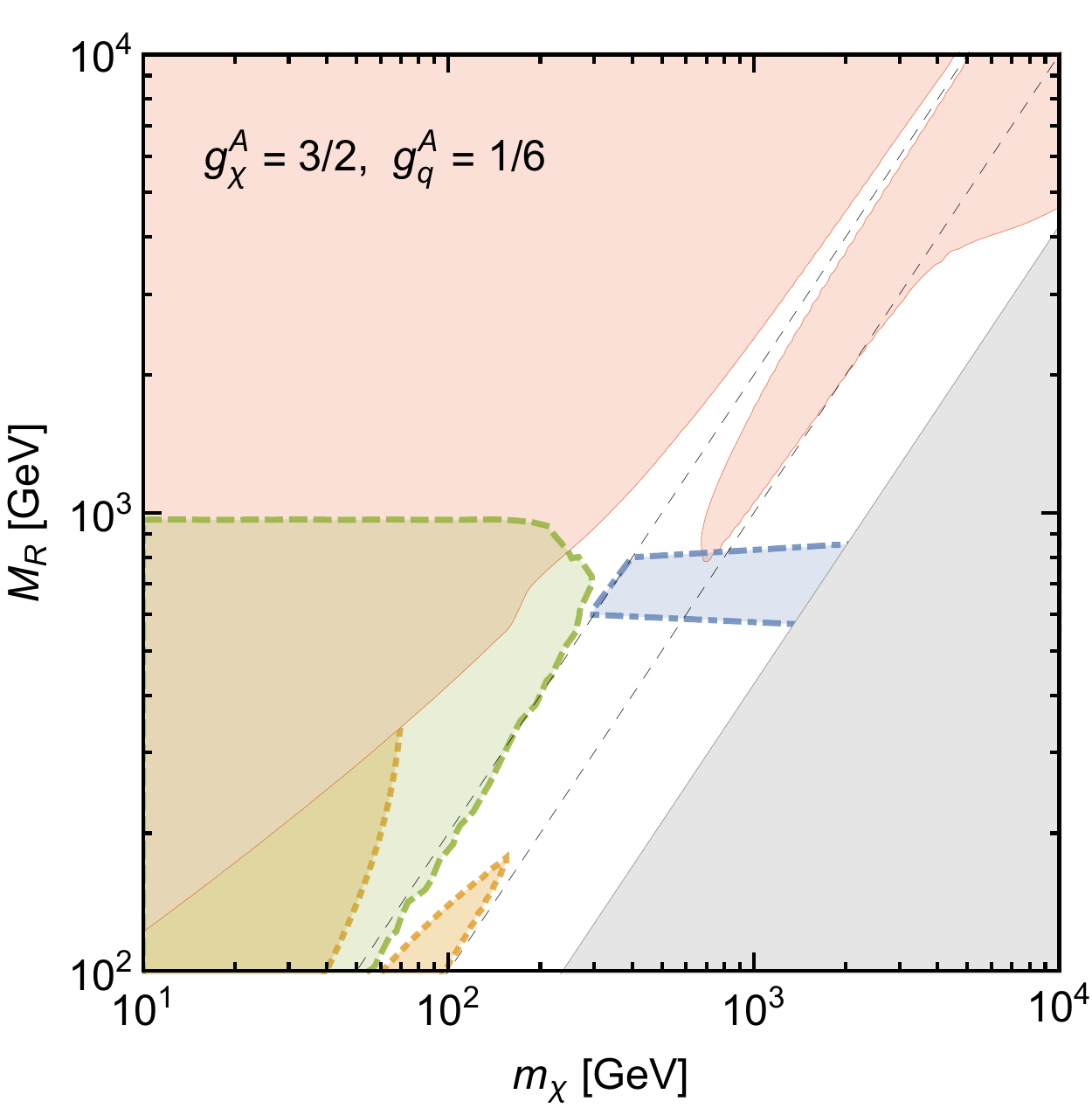}
\vspace{-2mm}
\caption{Combined constraints (at 95\% C.L.) from direct detection (orange, dotted), searches for monojets (green, dashed) and dijets (blue, dot-dashed) compared to the parameter region excluded by DSP overproduction (red) and perturbativity (grey). For the left (right) column, we have fixed $g \equiv (g^A_\chi \, g^A_q)^{1/2} = 1$ (g = 0.5), while the different rows show different coupling ratios $g^A_\chi / g^A_q$.}
\label{fig:universal}
\end{figure}

We are now in the position to combine the various constraints discussed in sections~\ref{sec:relic}--\ref{sec:dijet}. Our central results are shown in figures~\ref{fig:universal}--\ref{fig:vector}. In this section, we discuss these figures in detail and draw our conclusions on the complementarity of the different searches.

\subsection{Universal couplings}

In figure~\ref{fig:universal} we consider the case that the mediator couples to all quarks with equal coupling $g^A_q$, the different panels corresponding to different coupling strengths. For the left (right) column, we have fixed $g \equiv (g^A_\chi \, g^A_q)^{1/2} = 1$ (g = 0.5), while the different rows show different coupling ratios $g^A_\chi / g^A_q$ as discussed in section~\ref{sec:scenarios}. We make the following observations:
\begin{itemize}
 \item For $g = 1$, monojet searches are more strongly constraining than direct detection experiments for the entire parameter region where the DSP is underproduced in the early Universe (and the theory is perturbative). Considering coupling ratios \mbox{$g^A_\chi / g^A_q > 1$} further suppresses direct detection relative to monojet constraints, because the larger direct annihilation into pairs of mediators and the smaller mediator width reduces the DSP abundance.
 \item For $g = 0.5$, monojet searches have very limited sensitivity for the parameter region $m_\chi > M_R / 2$, corresponding to off-shell production of the DSP. Direct detection experiments therefore typically have an advantage over monojet searches in this region. Nevertheless, monojet searches are still clearly more constraining than direct detection for $m_\chi < M_R / 2$.
 \item For $g^A_\chi / g^A_q = 1$ and the coupling strengths under consideration essentially the entire mediator mass range $130\:\text{GeV} < M_R < (2.5\text{--}3)\:\text{TeV}$ is excluded by the dijet searches discussed in section~\ref{sec:dijet}. These constraints can be weakened by considering \mbox{$g^A_\chi / g^A_q > 1$}, both because the production cross section of the mediator is reduced and its invisible branching fraction is increased. In particular, for $g^A_\chi / g^A_q = 9$ dijet searches are insensitive to the parameter region $m_\chi < M_R / 2$, where the invisible branching fraction of the mediator is larger than 80\%. In this case, there is a very strong complementarity between monojet searches and dijet searches.
 \item For $g^A_q \geq 1/2$ the combination of the constraints from above excludes essentially the whole range of DSP and mediator masses apart from two special regions, namely $m_\chi \sim M_R / 2 \gtrsim 1 \: \text{TeV}$ and $m_\chi \gg M_R \sim M_Z$. For smaller quark couplings and large ratios $g^A_\chi / g^A_q$ additional allowed parameter regions open up for $m_\chi > M_R$.
\end{itemize}
For even smaller couplings than the ones considered in figure~\ref{fig:universal}, the bounds from the LHC can be significantly relaxed and we find that there is at present no sensitivity for $g^A_q, \, g^A_\chi < 1/6$. At the same time, the constraints from perturbativity are weakened, opening up additional parameter space for $m_\chi \gg M_R$. Nevertheless, direct detection constraints remain strong even for such small couplings and it becomes increasingly difficult to avoid an overproduction of the DSP. 

\subsection{Alternative coupling structures}

\begin{figure}[tb]
\centering
\includegraphics[width=0.4\textwidth, clip]{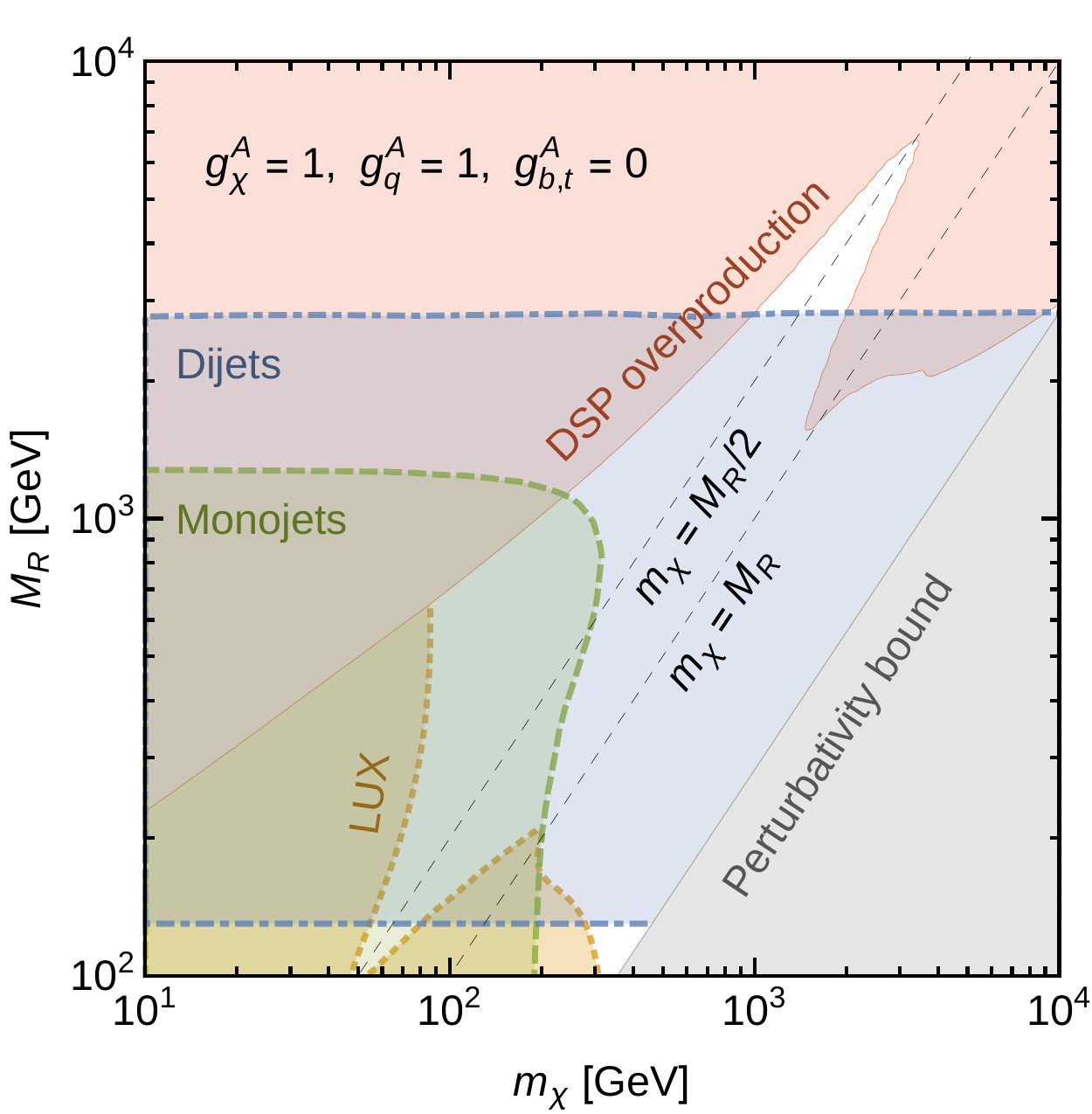}\quad
\includegraphics[width=0.4\textwidth, clip]{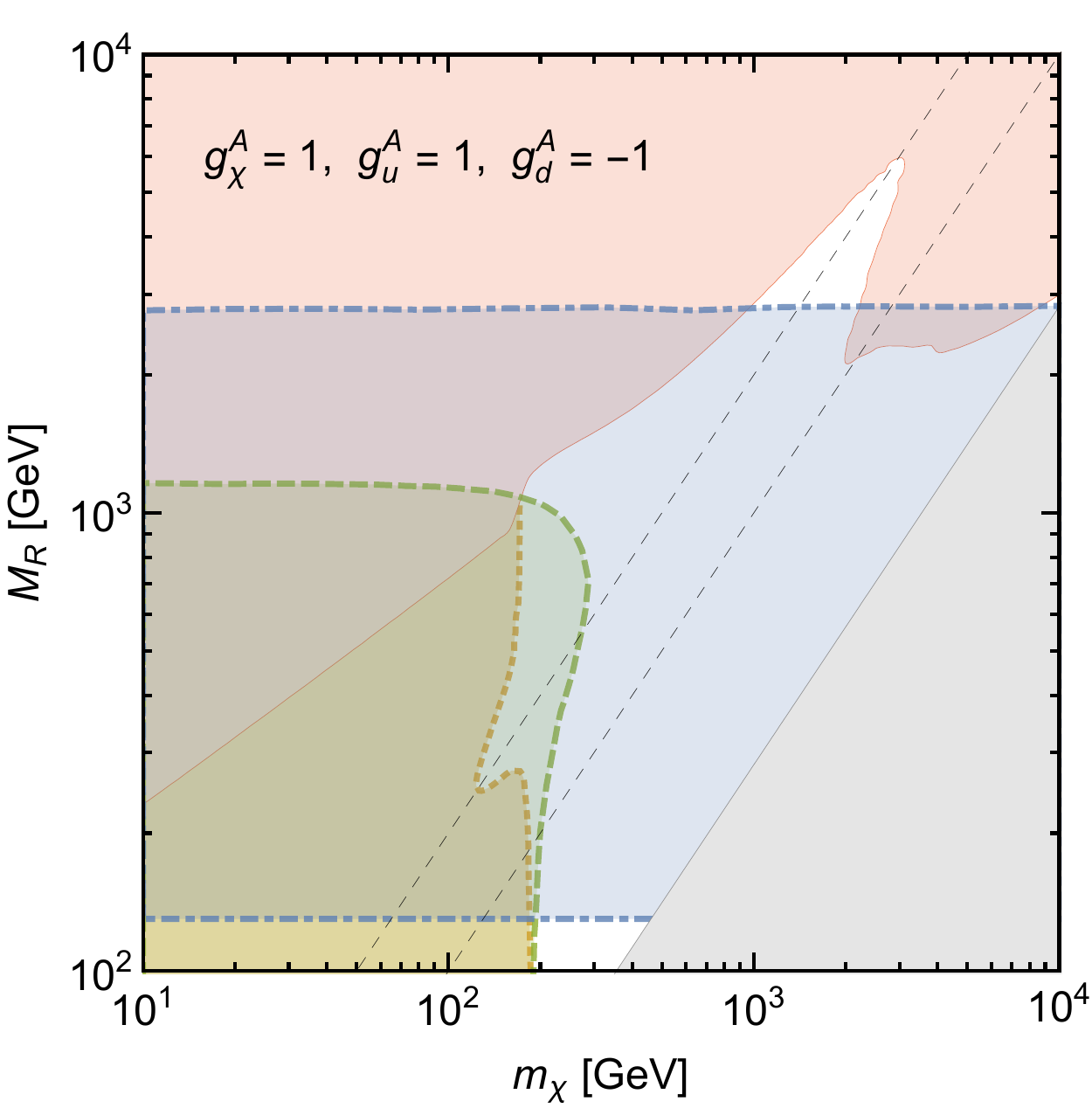}
\caption{Same as figure~\ref{fig:universal} but for alternative coupling structures. The left (right) plot shows the case of no heavy-quark couplings (isovector couplings).}
\label{fig:nonstandard}
\end{figure}

In figure~\ref{fig:nonstandard}, we consider two alternative scenarios for the coupling structure between the mediator and SM particles, namely couplings only to light quarks ($g^A_b = g^A_t = 0$) and isovector couplings ($g^A_u = -g^A_d$ etc.). In both cases, we have fixed $g^A_u = g^A_\chi = 1$.

The absence of couplings to heavy quarks changes the results compared to the case of universal couplings (see top-left panel of figure~\ref{fig:universal}) due to three separate effects:
\begin{enumerate}
 \item The most significant difference comes from the absence of the annihilation channel $\chi \bar{\chi} \rightarrow t \bar{t}$, which becomes important for $m_\chi > m_t$. As a result, the DSP abundance will be much larger than in the case of universal couplings and hence relic density constraints and direct detection constraints will become stronger.
 \item As a somewhat more subtle effect, the reduced number of decay channels means that the mediator will have a more narrow width (for fixed coupling strength), leading to a smaller relic density and hence weaker constraints from the relic density requirement and direct detection in the resonance region $m_\chi \sim M_R / 2$.
 \item For the same reason, the mediator will have a larger invisible branching fraction, enhancing the constraints from monojet searches in the parameter region $m_\chi < M_R / 2$ (see also figure~\ref{fig:monojets}).
\end{enumerate}
We find that as a result of all of these effects direct detection is enhanced relative to LHC searches, although the combined bound is
still dominated by the LHC for $g = 1$.

The sign difference for isovector couplings will not change LHC cross sections, decay widths and relic density calculations, all of which are independent of interference effects.\footnote{One notable exception are DM searches in the mono-$W$ channel, where the relative sign between up-quark and down-quark coupling plays an important role~\cite{Bai:2012xg}. For isovector couplings, these searches are expected to give stronger bounds than conventional monojet searches (see e.g.~\cite{Aad:2013oja}), but a detailed analysis in the context of our model is beyond the scope of the present work.} However, interference between the individual quark contributions is crucial for direct detection. One can immediately infer from the values for $\Delta q^{(N)}$ given in eq.~(\ref{eq:deltas}) that there is destructive interference between the up-quark and the down-quark (as well as the strange-quark) contribution if all couplings have equal signs. For isovector couplings, on the other hand, the interference is constructive. As a result, this simple modification of the standard scenario increases the effective DSP-nucleon coupling by a factor of 3 and hence boosts direct detection cross sections by an order of magnitude. In spite of this enhancement, however, we still find LHC searches to more constraining than direct detection even for isovector couplings.

\subsection{Vector couplings}
\label{sec:vector}

\begin{figure}[tb]
\centering
\includegraphics[width=0.42\textwidth, clip]{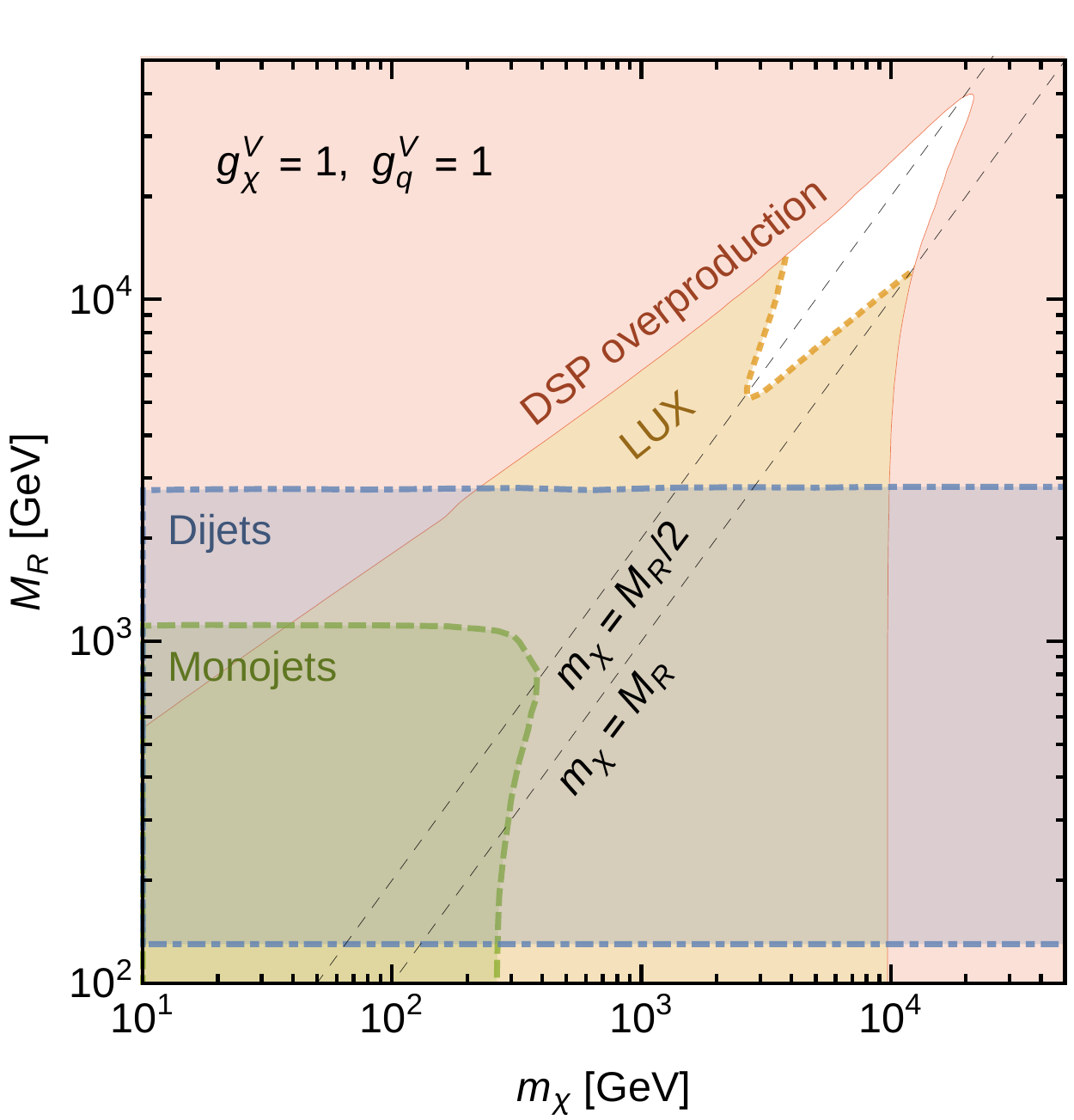}\quad
\includegraphics[width=0.43\textwidth, clip]{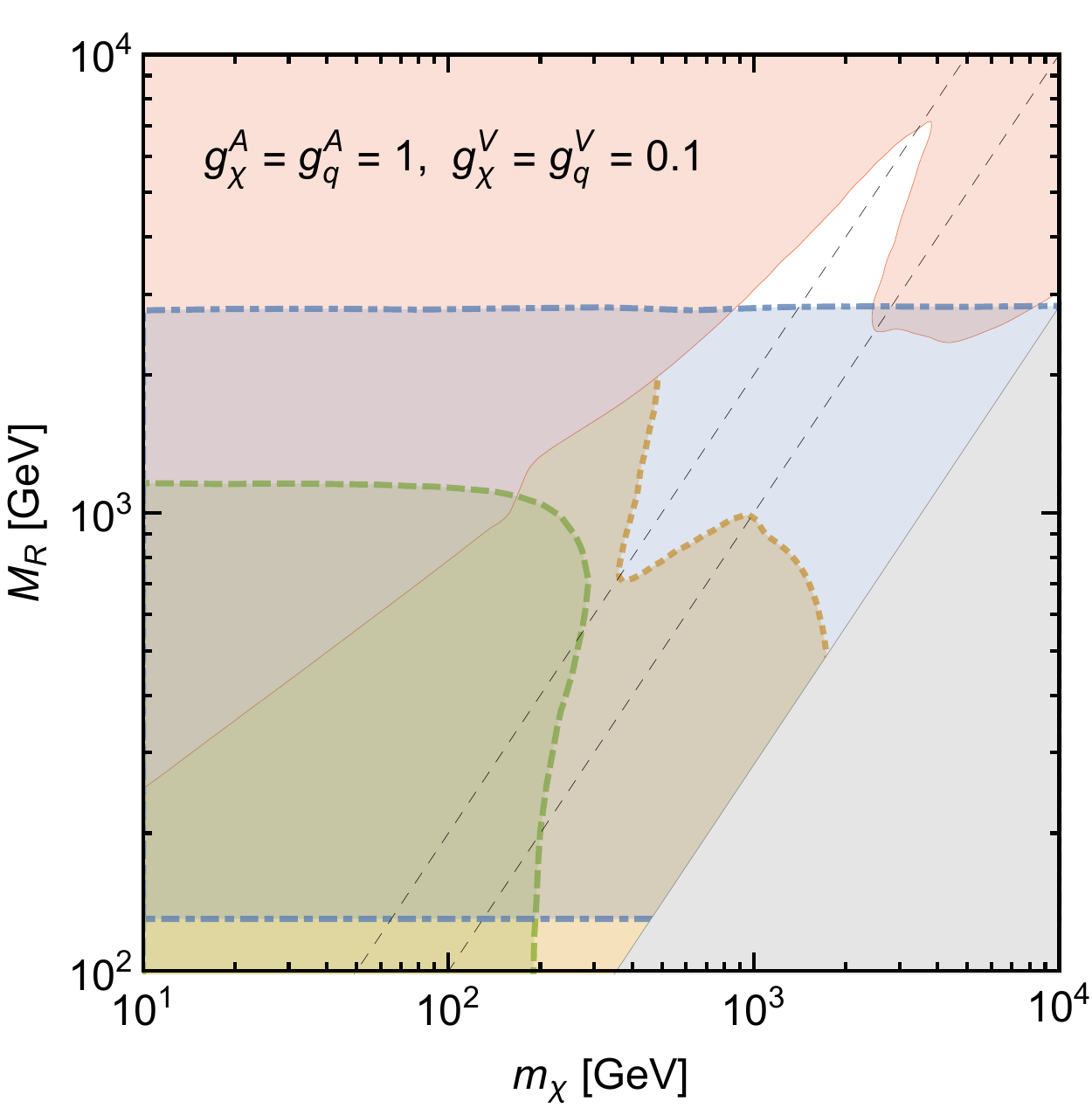}
\caption{Same as figure~\ref{fig:universal} but in the presence of vector couplings. The left (right) plot shows the case where $g^V \gg g^A$ ($g^A \gg g^V$). Note the change of scale in the left figure.}
\label{fig:vector}
\end{figure}

To conclude this section, let us briefly consider non-zero vector couplings $g^V_q$ and $g^V_\chi$. Figure~\ref{fig:vector} (left) shows the combined bounds for $g^A_q = g^A_\chi = 0$ and $g^V_q = g^V_\chi = 1$. As expected, direct detection receives a huge enhancement due to the presence of SI interactions. As a result, LHC searches are essentially irrelevant in comparison. The only exception is the parameter region $m_\chi \lesssim 10\:\text{GeV}$ (not shown in figure~\ref{fig:vector}), where direct detection loses sensitivity. This region, however, is already fully excluded by the combination of monojet searches and the relic density requirement. 

As a somewhat more interesting case, let us consider the case $g^A_q, \, g^A_\chi \gg g^V_q, \, g^V_\chi > 0$. For such a setup, the vector couplings give no relevant contribution to the total width of the mediator and they play no role for the LHC phenomenology of the model. Nevertheless, small vector couplings can potentially give a relevant contribution to the relic density calculation (due to the absence of a helicity suppression in the $s$-wave contribution) and they will certainly lead to an enhancement of event rates in direct detection experiments. One particular example is shown in figure~\ref{fig:vector} (right). For $g^A_q = g^A_\chi = 1$ and $g^V_q = g^V_\chi = 0.1$ direct detection experiments clearly give stronger bounds than monojet searches, but they are less constraining for heavy mediator masses than dijet searches.

\section{Discussion}
\label{sec:discussion}

As we have seen in the previous section, the parameter space under consideration can be tightly constrained by combining results from a variety of colliders searches. This observation immediately leads to two important questions: First, can the remaining parameter space be constrained even further using near-future experiments? And second, can these constraints be evaded by modifying the assumptions on the coupling structure and the cosmology of the dark sector? We will now address both of these questions in turn.

\subsection{Future prospects}

Within the next decade we can expect rapidly increasing target masses and exposures in direct detection experiments as well as a significant improvement in the centre-of-mass energy and luminosity at the LHC. To discuss the expected impact of these experimental developments, we consider two different projections:
\begin{enumerate}
 \item A three-year projection considering the LHC with an integrated luminosity of $100\:\text{fb}^{-1}$ at 14 TeV and a direct detection experiment based on liquid xenon with a total exposure of 2 ton-years (like XENON1T~\cite{XENON1T}).
 \item A ten-year projection considering the LHC with an integrated luminosity of $300\:\text{fb}^{-1}$ at 14 TeV and a direct detection experiment based on liquid xenon with a total exposure of 10 ton-years (like LZ~\cite{LZ}).
\end{enumerate}

For the projected sensitivity of direct detection experiments we assume that these experiments will remain essentially free of backgrounds so that the resulting bounds on the DSP-nucleon scattering cross section is inversely proportional to the total exposure. XENON1T can hence improve the current bound from LUX by a factor of 70, while LZ can achieve a factor of 350 improvement.\footnote{As long as we consider only DSP masses above $10\:\text{GeV}$, the sensitivity of LZ is not limited by the background from solar neutrinos~\cite{Ruppin:2014bra}.}

In~\cite{Monojets}, the ATLAS collaboration has studied the sensitivity of the LHC at 14 TeV for monojets. We have generated sets of monojet events based on the cuts suggested there, imposing in particular a very stringent cut on missing transverse energy of $\slashed{E}_T > 800\:\text{GeV}$. For these cuts, the expected bounds on the fiducial monojet cross section are around $1\:\text{fb}$.\footnote{The precise numbers can be worked out by comparing the numbers of predicted events given in table~3 with the expected bounds shown in figure~6 of~\cite{Monojets}.} It is important to note that while a huge improvement in sensitivity will be achieved from the larger centre-of-mass energy, which better separates signal from background, the gain from growing luminosity is rather slow due to significant systematic uncertainties in the background estimation (assumed to be 5\%).

\begin{figure}[tb]
\centering
\includegraphics[width=0.42\textwidth, clip]{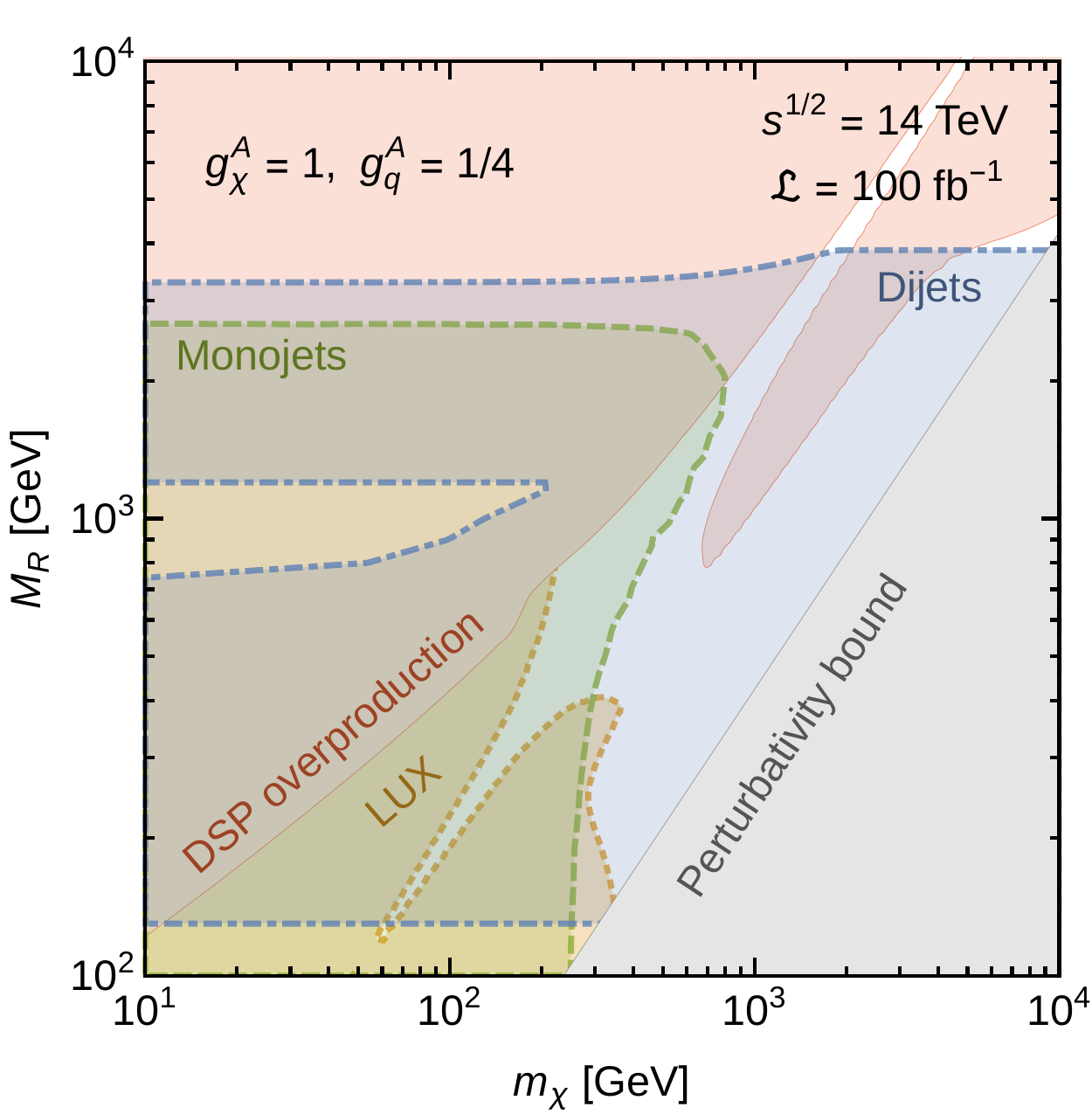}\quad
\includegraphics[width=0.42\textwidth, clip]{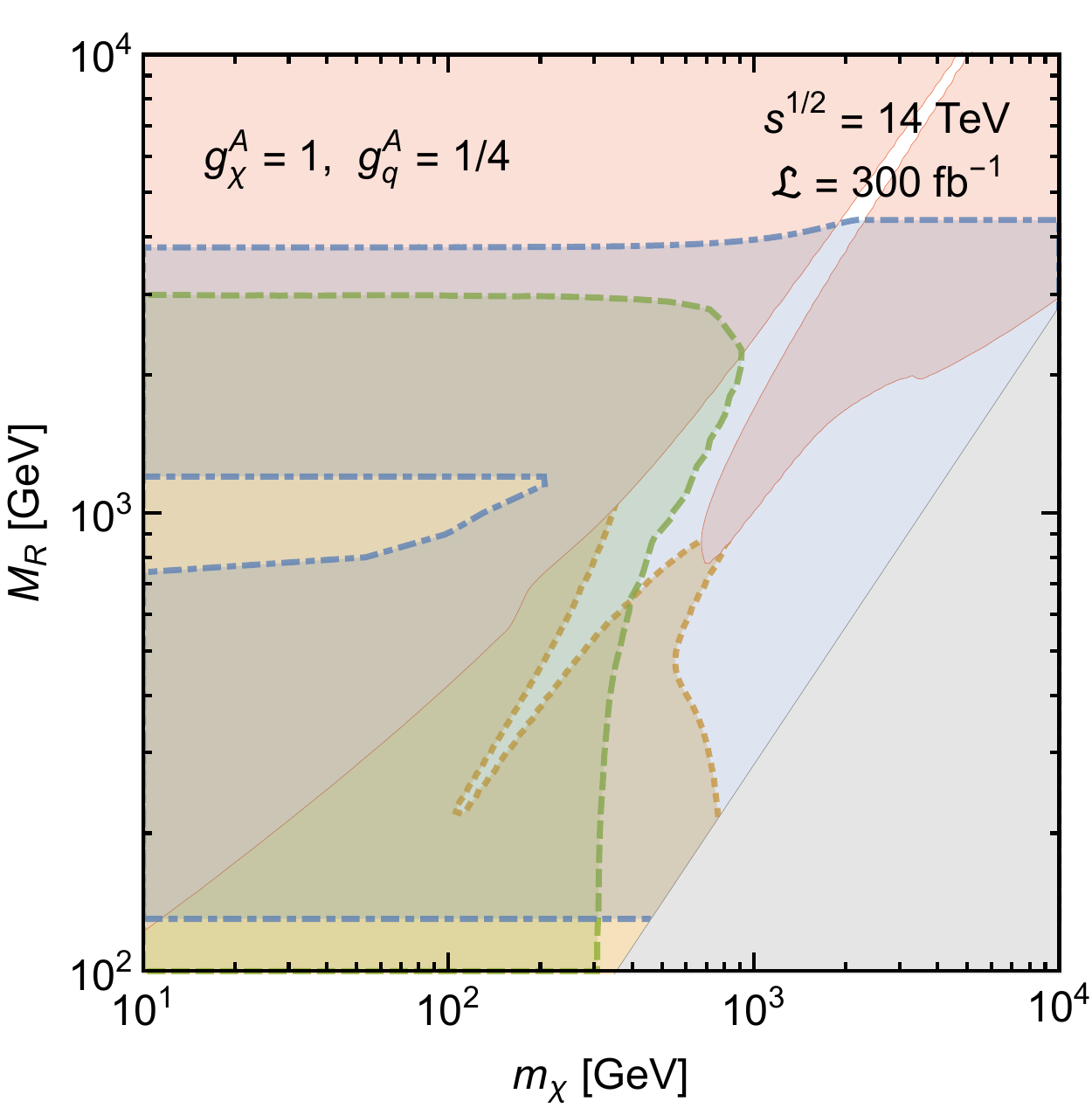}
\caption{Same as figure~\ref{fig:universal} but showing projected sensitivities for near-future experiments. We focus on the case $g^A_q = 1/4$ and $g^A_\chi = 1$ (cf.~centre-right panel of figure~\ref{fig:universal}) and show a three-year (ten-year) projection on the left (right). See text for details.}
\label{fig:projection}
\end{figure}

Various previous works have discussed the projected sensitivity of dijet searches at future colliders. Ref.~\cite{Chiang:2015ika}, for example, has considered prospects for searches for dijet resonances in association with SM gauge bosons. The best expected sensitivity was found for the case where the gauge boson is a photon. We take the projected bounds for the one-Higgs doublet case (called BP1 in~\cite{Chiang:2015ika}), having confirmed explicitly that the production cross section of the mediator is the same for vector and axial couplings. A dedicated study of dijet resonances at higher energies has been performed in~\cite{Yu:2013wta}. We adopt their bound for an integrated luminosity of $300\:\text{fb}^{-1}$ and infer the corresponding bound for $\mathcal{L} = 100\:\text{fb}^{-1}$ by using that bounds on the coupling $g_q$ scale with luminosity approximately proportional to $\mathcal{L}^{1/4}$, because bounds on the production cross section should become stronger proportional to $\sqrt{\mathcal{L}}$ and the production cross section is proportional to $g_q^2$.\footnote{Ref.~\cite{Yu:2013wta} considers only the case of very narrow resonances. Nevertheless, we have confirmed explicitly that one obtains comparable projections for $\Gamma_R / M_R \approx 0.03\text{--}0.05$, corresponding to the couplings considered in figure~\ref{fig:projection}.}

We note that (projected) constraints on dijet resonances are typically shown as constraints on the mediator-quark coupling, i.e. $g_q < g_\text{max}(M_R)$, assuming that the mediator can only decay into quarks. In the presence of an additional invisible decay channel, these constraints can be considerably weakened. To apply constraints from the literature to our model, we therefore have to rescale any bound according to\footnote{An additional rescaling factor of $1/6$ arises when comparing with~\cite{Dobrescu:2013cmh,Yu:2013wta} due to different conventions for the couplings.}
\begin{equation}
 g^\text{rescaled}_\text{max}(M_R, m_\chi, g_\chi) = \sqrt{\frac{\text{BR}(R \rightarrow jj)_\text{quarks-only}}{\text{BR}(R \rightarrow jj)_\text{quarks+DSP}}} g_\text{max}(M_R) \;.
\end{equation}

Presently there are no sensitivity studies for the intermediate mass region $600\:\text{GeV} < M_R < 1200\:\text{GeV}$ and, in light of the large resources required to deal with the huge statistics in this mass range in dijet searches, it is not clear whether future colliders with higher centre-of-mass energy can significantly improve their sensitivity. An interesting option~\cite{Doglioni} to overcome this difficulty consists of recording either a fraction of the events (this procedure has been already applied in~\cite{Aad:2014aqa}) or a reduced part of each event (see e.g.~\cite{Dijets}). {A different possibility} to make progress may be searches for $t \bar{t}$ resonances along the lines of~\cite{Aad:2013nca,Chatrchyan:2012yca}, which would also be a generic signature of the model considered here. We leave a detailed study of these searches for future work and simply use existing bounds from the Tevatron and the LHC for the intermediate mass region.

In figure~\ref{fig:projection} we consider the case $g^A_q = 1/4$, $g^A_\chi = 1$, which is comparably weakly constrained by current experiments (see centre-right panel of figure~\ref{fig:universal}). We can see direct detection experiments rapidly gaining sensitivity in the parameter region $m_\chi > M_R / 2$, while monojet searches can probe the resonance region $m_\chi \sim m_R / 2$ up to mediator masses of around $2\:\text{TeV}$. Finally, dijet searches will be sensitive to values of $M_R$ up to $(3\text{--}4)\:\text{TeV}$. We note that a possible future $100\:\text{TeV}$ hadron collider could easily reach up to $M_R \sim 10\text{--}20\:\text{TeV}$, placing extremely strong limits on candidate dark sector theories~\cite{100tev, Xiang:2015lfa, Yu:2013wta}.

\subsection{Generalised bounds}

Let us now turn to a discussion of the model assumptions that we have made. Crucially, all of our results have been based on the assumption that the width of the mediator is dominated by the couplings between the mediator and quarks as well as DSPs. Similarly, we have assumed that only these couplings are relevant for the relic density calculation. While it is justified to neglect additional couplings to leptons and SM gauge bosons given the stringent experimental bounds, it is of course conceivable that there are additional unstable states in the dark sector, which are light enough to provide additional channels for DSP annihilation and/or mediator decay. If these additional light states subsequently decay into high-multiplicity SM states, it is conceivable that they could evade detection in all existing collider searches. The presence of such additional states implies that we can no longer calculate $\Gamma_R$ and $\Omega_\text{DSP}$ in terms of the couplings $g_q$ and $g_\chi$ and have to treat them as additional free parameters.\footnote{In principle, the presence of additional light states can only lead to an increase in $\Gamma_R$ and a decrease in $\Omega_\text{DSP}$, so that these parameters cannot be chosen arbitrarily for given $g_q$ and $g_\chi$. However, we neglect this complication for the purpose of the present discussion.} 

\begin{figure}[tb]
\centering
\includegraphics[width=0.42\textwidth, clip]{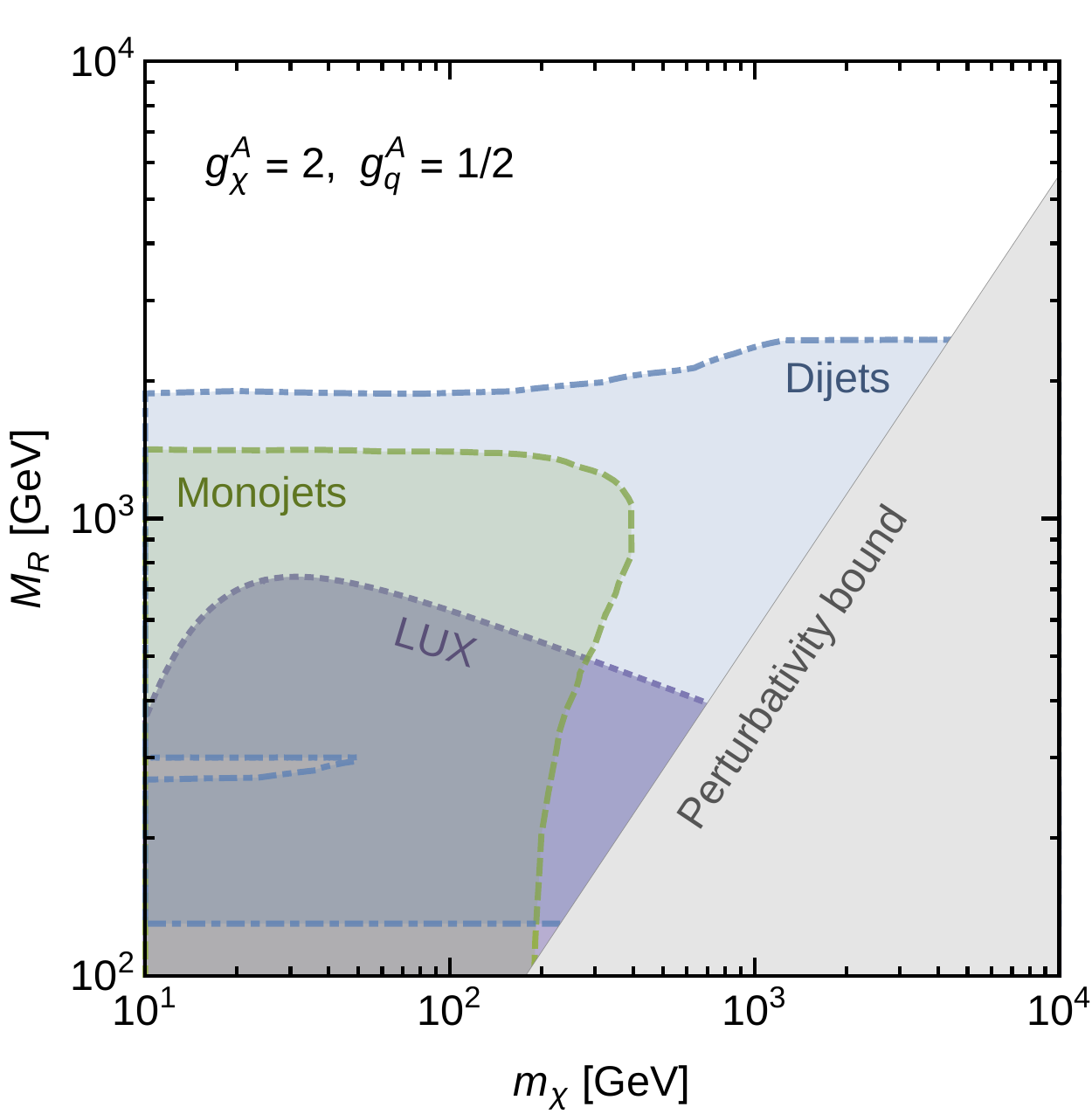}\quad
\includegraphics[width=0.42\textwidth, clip]{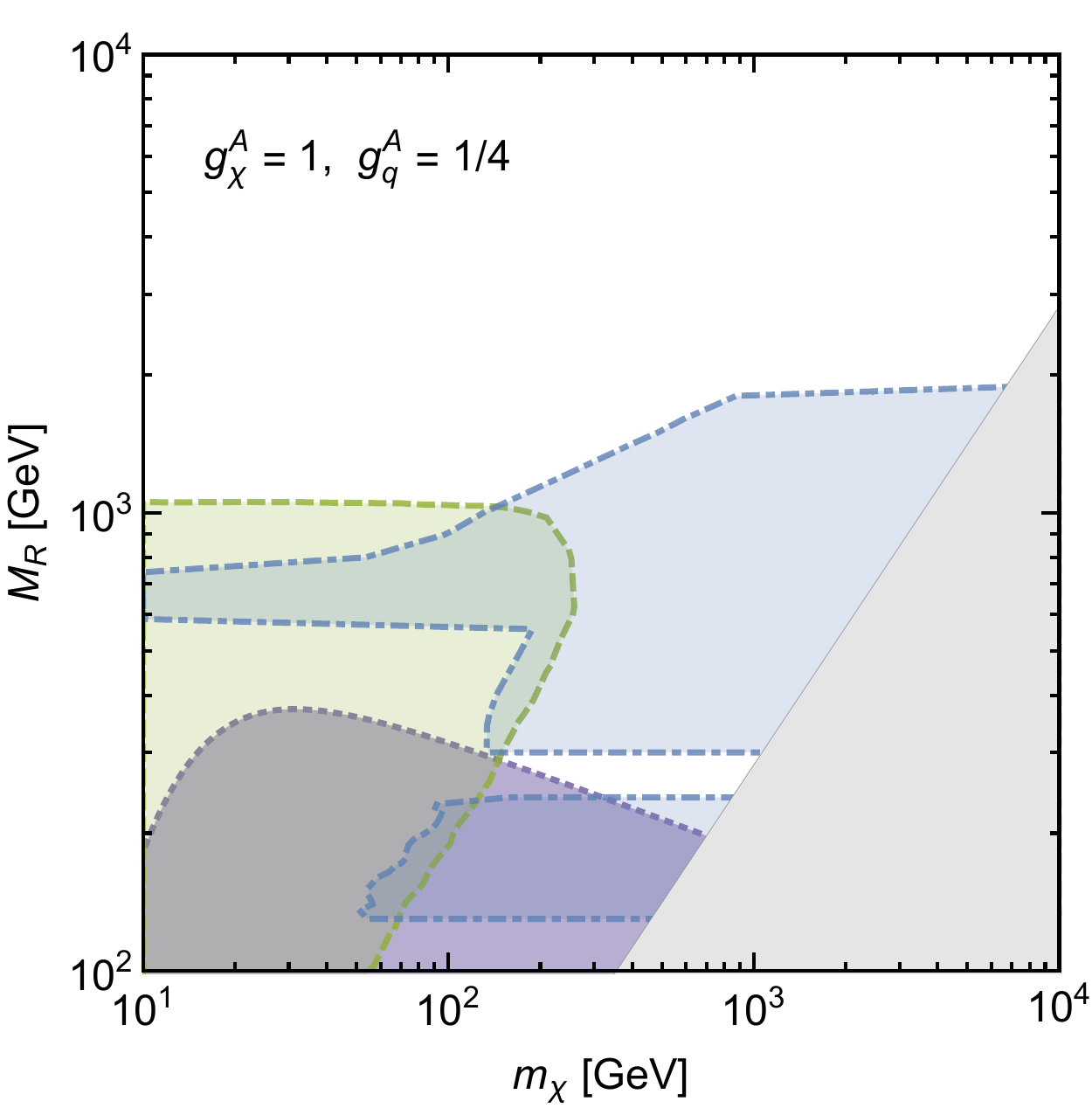}
\caption{Same as figure~\ref{fig:universal} but fixing the DSP relic abundance $\Omega_\text{DSP} = \Omega_\text{DM}$. As a result, there is no longer an excluded parameter region corresponding to DSP overproduction and the constraint from direct detection is significantly strengthened in the parameter region where previously the DSP was underproduced. To make this change of perspective explicit, we show the direct detection bound in purple rather than orange (cf.~figure~\ref{fig:DD}).}
\label{fig:alternativerelic}
\end{figure}

In figure~\ref{fig:alternativerelic} we show two examples for how our bounds would change if we fixed $\Omega_\text{DSP} = \Omega_\text{DM}$ rather than calculating the relic density in terms of $g_q$ and $g_\chi$. By construction, there is no longer an excluded parameter region corresponding to the overproduction of the DSP, so the parameter region $M_R \gg 1\:\text{TeV} \gg m_\chi$ is no longer excluded. At the same time, bounds from direct detection are significantly strengthened in the parameter region where previously the DSP was underproduced. Consequently, these bounds are now more competitive with monojet and dijet searches, which are not affected by changing the value of $\Omega_\text{DSP}$.

In contrast, increasing the total width of the mediator will strongly affect dijet searches, because a broader resonance will be less visible in the dijet invariant-mass distribution and furthermore the presence of additional decay channels will reduce the branching ratio of the mediator into light quarks. The modified branching ratios also imply that there will be weaker bounds from monojet searches in the parameter region where the mediator can decay into DSPs. For heavy DSP masses, on the other hand, the monojet cross section is largely independent of the mediator width (cf.~section~\ref{sec:monojet}).

\begin{figure}[tb]
\centering
\includegraphics[width=0.48\textwidth, clip]{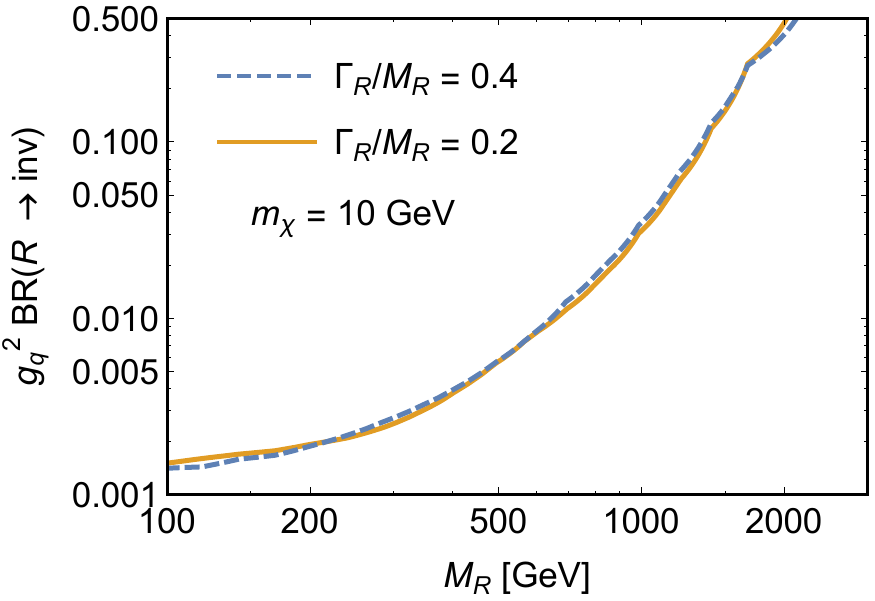}\quad
\includegraphics[width=0.48\textwidth, clip]{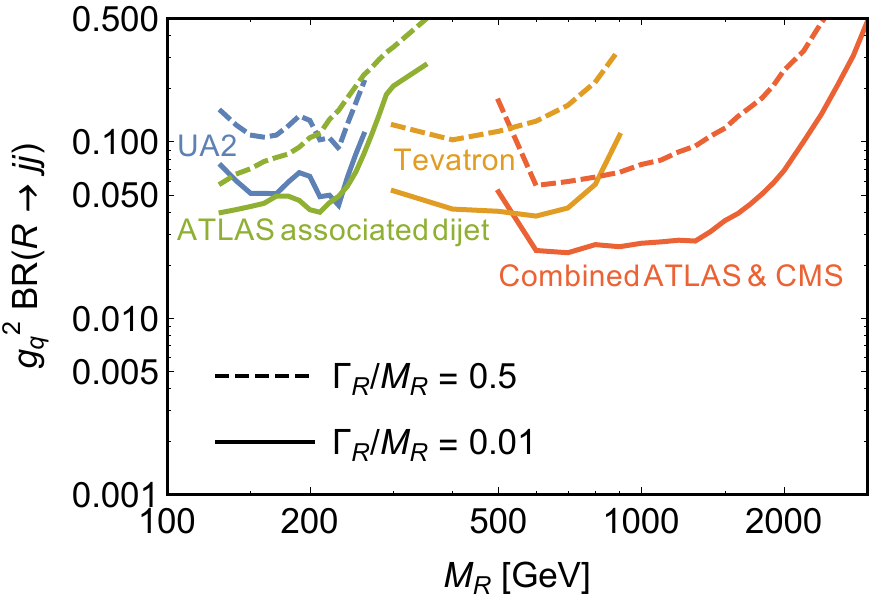}
\caption{Experimental bounds on $g_q^2 \times \text{BR}(R \rightarrow \text{inv})$ (left) and $g_q^2 \times \text{BR}(R \rightarrow jj)$ (right) as a function of $M_R$ for different values of $\Gamma_R$.}
\label{fig:alternativewidths}
\end{figure}

To study the dependence of the monojet and dijet bounds on the mediator width in more detail, we show in figure~\ref{fig:alternativewidths} the experimental bounds on $g_q^2 \times \text{BR}(R \rightarrow \text{inv})$ (left panel) and $g_q^2 \times \text{BR}(R \rightarrow jj)$ (right panel) for different values of $\Gamma_R$.\footnote{Note that these bounds apply for axial couplings as well as vector couplings. If both kinds of couplings are non-zero, the bounds apply on the combination $g_q^2 = \left(g^A_q\right)^2 + \left(g^V_q\right)^2$.} The combination $g_q^2 \times \text{BR}$ is chosen because~--- within the validity of the narrow-width approximation~--- the resulting bound is expected to be independent of the mediator width. Indeed, the left panel of figure~\ref{fig:alternativewidths} confirms explicitly that for $m_\chi \ll M_R$ the monojet cross section is proportional to $g_q^2 \times \text{BR}(R \rightarrow \text{inv})$ even for rather broad resonances, i.e.\ the dependence on the width of the mediator only enters via the branching ratios. 

In the right panel of figure~\ref{fig:alternativewidths}, on the other hand, we can clearly see the loss of sensitivity of dijet searches for broad resonances. As the mediator width is increased, the bound on $g_q^2 \times \text{BR}(R \rightarrow jj)$ becomes weaker. 
Note that, in the present work we do not consider resonances with a width larger than 50\% of its mass. For such resonances it is very difficult to distinguish the dijet invariant-mass distribution of the signal from backgrounds. A more promising strategy to constrain very broad resonances might therefore be to study dijet angular correlations~\cite{deVries:2014apa}, such as the ones considered in the context of constraining four-fermion operators (see e.g.~\cite{Khachatryan:2014cja}). We leave this line of investigation for future work.

To conclude, we emphasise that the bounds shown in figure~\ref{fig:alternativewidths} can in principle be applied to any model containing a spin-1 resonance that can decay into quarks and invisible particles (where the latter could also be SM neutrinos). For a given mass of the resonance, one simply needs to calculate the total width of the resonance and the respective branching ratios and then compare the model prediction with the appropriate bounds. The specific model of an axial-vector mediator, which we have chosen to present our results, therefore only serves for the purpose of illustration, but does not limit the applicability of our analysis.

\section{Conclusions}
\label{sec:conclusions}

In this work, we have discussed the phenomenology of a dark sector particle (DSP) with relatively large couplings to quarks. We considered an axial-vector mediator as a suitable example to illustrate the compelling interplay between different DM detection techniques, as well as the impact of additional theoretical considerations. The relevant constraints arise from collider searches for missing energy and resonances, DM direct detection experiments, the DSP relic abundance and perturbativity, each of which has a unique foothold in the search for DSPs.

To calculate constraints from direct detection experiments, we
calculate the DSP relic abundance $\Omega_\text{DSP}$ based on
the assumed interactions and account for the fact that the local DSP
density is proportional to $\Omega_\text{DSP}$. This approach suppresses constraints from
direct detection experiments in the parameter region where the DSP is
underproduced. Moreover, this approach implies that direct detection
bounds are largely invariant under a rescaling of couplings (see
figure~\ref{fig:DD}), making them less sensitive than LHC searches whenever
the couplings between the DSP and quarks are large.

Two classes of LHC constraints are important in the present context: Searches for DSP production in events with large amounts of missing transverse energy in association with SM particles (e.g. monojet events) and direct constraints on the mediator from resonance searches. While the former have been intensely studied in the literature, the latter have received much less attention in the context of DM searches. In fact, it has often been assumed that these kinds of searches give only very weak constraints on mediators with small mass or large width.

In the present work we provide a comprehensive analysis of searches for dijet resonances in UA2, the Tevatron and the LHC across a wide range of mediator masses. In particular, we reinterpret an existing ATLAS analysis for dijet resonances produced in association with leptonically decaying gauge bosons in the context of our model and obtain strong bounds in the region where the new mediator $R$ is rather light, namely $130\:\text{GeV} \leq M_R \leq 300\:\text{GeV}$. Furthermore, we show that, although the experimental searches become less constraining for broader resonances, there are still stringent bounds on mediators even with
width $\Gamma_R \sim M_R/2$ (see figure~\ref{fig:dijets}). These bounds need to be taken into account when interpreting DM searches at the LHC in terms of simplified models with an $s$-channel mediator, because they apply to a wide range of models and in many cases complement or even surpass other search strategies (see figure~\ref{fig:alternativewidths}).

For example, for the choice of (axial) couplings of $R$ to quarks and DSPs \mbox{$g^A_q = g^A_\chi = 1$} all mediator masses in the range $130\:\text{GeV} \lesssim M_R \lesssim 3\:\text{TeV}$ are excluded by dijet searches, independent of the DSP mass. Only two small parameter regions remain viable, namely the resonance region $M_R \approx 2 \, m_\chi \gg 1\:\text{TeV}$ and the low-mass region $m_\chi \gg M_R \approx M_Z$, where the DSP annihilates directly into the mediator (see figure~\ref{fig:universal}). These constraints can be significantly weakened by considering $g^A_\chi / g^A_q \gg 1$ and we have studied the cases $g^A_\chi / g^A_q = 4$ and $g^A_\chi / g^A_q = 9$ in detail. For example, for $g^A_q = 1/4$ and $g^A_\chi = 1$ there is a strong complementarity between monojet and dijet searches, but there are still significant mass ranges unconstrained by present data. These couplings thus provide an interesting benchmark point to study projections for near-future experiments (see figure~\ref{fig:projection}).

We have furthermore pointed out that in the presence of non-zero axial couplings, it is not possible within a simple perturbative theory to raise the DSP mass $m_\chi$ arbitrarily above the mediator mass $M_R$. The same problem also manifests itself in the annihilation process $\chi \bar{\chi} \rightarrow t \bar{t}$ violating perturbative unitarity for large DSP masses. This observation implies that a significant part of the $M_R$-$m_\chi$ parameter plane is theoretically inaccessible. In combination with the constraints on the mediator mass from the relic density calculation, we obtain upper bounds on both the mediator mass and the DSP mass for fixed couplings, so that the relevant parameter space is necessarily closed.

With improving sensitivity at direct detection experiments and colliders it will be possible to probe the allowed mass ranges for smaller and smaller couplings, up to the point where DSP annihilation into quarks will generally be insufficient to avoid DSP overproduction. In other words, there is the realistic chance to comprehensively test the idea that a DSP can have large interactions with SM quarks mediated by a spin-1 particle. Such a conclusion will not be achieved by a single experiment, but it will require significant progress across a range of different experimental strategies combined with continuing theoretical studies of less-explored search channels.

\acknowledgments

We thank Nathaniel Craig, Mads Frandsen, Ulrich Haisch, Sungwon Lee and Ian Shoemaker for helpful discussions. We are grateful to Juan P.~Araque for providing help with \texttt{MCLimit} as well as to {Caterina Doglioni, Matthew Dolan, Christopher McCabe and David Salek} for valuable comments on the manuscript. FK would like to thank the Munich Institute for Astro- and Particle Physics for hospitality during the workshop \emph{Dark MALT 2015}, where part of this work was carried out. This work is supported by the German Science Foundation (DFG) under the Collaborative Research Center (SFB) 676 Particles, Strings and the Early Universe. 

\providecommand{\href}[2]{#2}\begingroup\raggedright\endgroup

\end{document}